\pdfoutput=1
\documentclass[aps,groupaddress,superscriptaddress,reprint,amsmath,amssymb,graphicx
]{revtex4-1}

\usepackage{amsmath}
\usepackage{amsfonts}
\usepackage{amssymb}

\usepackage{color}
\usepackage{graphicx}

\usepackage{hyperref}
\usepackage{enumerate}

\usepackage{comment}

\usepackage{mathrsfs}
\usepackage{float}
\usepackage[T1]{fontenc}

\begin{document}

\title{Fluctuations of entropy production of a run-and-tumble particle}

\author{Prajwal Padmanabha}
\affiliation{Department of Physics and Astronomy ``G. Galilei'', University of Padova, Padova 35131, Italy}
\author{Daniel Maria Busiello}
\thanks{current address: Max Planck Institute for the Physics of Complex Systems, 01187 Dresden, Germany}
\affiliation{Ecole Polytechnique F\'{e}d\'{e}rale de Lausanne (EPFL), 1015 Lausanne, Switzerland}
\author{Amos Maritan}
\affiliation{Department of Physics and Astronomy ``G. Galilei'', University of Padova, Padova 35131, Italy}
\author{Deepak Gupta}
\affiliation{Department of Physics, Simon Fraser University, Burnaby, British Columbia V5A 1S6, Canada}
\affiliation{Institute for Theoretical Physics, Technical University of Berlin, Hardenbergstr. 36, D-10623 Berlin, Germany}
\begin{abstract}
Out-of-equilibrium systems continuously generate entropy, with its rate of production being a fingerprint of non-equilibrium conditions. In small-scale dissipative systems subject to thermal noise, fluctuations of entropy production are significant. Hitherto, mean and variance have been abundantly studied, even if higher moments might be important to fully characterize the system of interest. Here, we introduce a graphical method to compute any moment of entropy production for a generic discrete-state system. Then, we focus on a paradigmatic model of active particles, i.e., run-and-tumble dynamics, which resembles the motion observed in several microorganisms. Employing our framework, we compute the first three cumulants of the entropy production for a discrete version of this model. We also compare our analytical results with numerical simulations. 
We find that as the number of states increases, the distribution of entropy production deviates from a Gaussian. Finally, we extend our framework to a continuous state-space run-and-tumble model, using an appropriate scaling of the transition rates. The approach here presented might help uncover the features of non-equilibrium fluctuations of any current in biological systems operating out-of-equilibrium.
\end{abstract}

\maketitle

\section{Introduction}
Biological systems are often found out of equilibrium, constantly consuming energy to maintain a stationary state \cite{prigogine1971biological,Schrodinger1944}. A large number of studies have been performed on non-equilibrium properties that such systems display \cite{fang2019nonequilibrium,ritort2008nonequilibrium,bustamante2005nonequilibrium}. One of the most relevant signatures of a non-equilibrium condition is the net production of entropy. At equilibrium, due to the time-reversal symmetry, a forward trajectory is equally probable compared to its time-reversed counterpart. Non-equilibrium conditions break this symmetry, leading to entropy production. This has been closely investigated, especially in the context of biological systems, to quantify their distance from thermodynamic equilibrium \cite{fodor2016far,li2019quantifying}. Even outside the context of biological systems, entropy production and its features have been extensively studied \cite{seifert2012stochastic, diana2014mutual, busiello2017entropy, busiello2019entropy, busiello2019entropyB}. 

Moreover, investigating macroscopic emergent behaviors is usually not enough in the realms of biological and biochemical systems. In fact, thermal fluctuations are prominent in small-scale systems. Therefore, researchers have investigated fluctuations of different thermodynamic quantities, such as work done \cite{van2003stationary,douarche2006work,sabhapandit2012heat}, entropy production~\cite{wang2002experimental,ciliberto2013heat,gupta2018partial,shreshtha2019thermodynamic}, heat flow~\cite{Van-Zon-heat,visco2006work,sabhapandit2012heat,kundu2011large}, and stochastic efficiency~\cite{martinez2016brownian,van2012efficiency,verley2014unlikely,gupta2017stochastic,gupta2018exact} within the framework of stochastic thermodynamics \cite{seifert2005entropy,seifert2012stochastic,sekimoto2010stochastic}.

In stark contrast to equilibrium systems, basic principles are constantly being searched for in  non-equilibrium systems. Stochastic thermodynamics provides a window into the possibilities of out-of-equilibrium universal laws through several seminal results, such as the fluctuation theorems \cite{kurchan1998fluctuation,crooks1999entropy,lebowitz1999gallavotti,campisi2009fluctuation}, the Jarzynski equality \cite{jarzynski1997nonequilibrium}, the Crooks work-fluctuation theorem \cite{crooks1999entropy,crooks1998nonequilibrium}, the non-equilibrium linear response \cite{baiesi2013update,baiesi2009fluctuations}, and the thermodynamic uncertainty relations \cite{dechant2018multidimensional,barato2015thermodynamic,gingrich2016dissipation,horowitz2020thermodynamic}.

In particular, entropy production and its fluctuations play a prominent role in linear response theory, fluctuation theorems, and thermodynamic uncertainty relations. In this context, various studies have focused on the estimation of the mean entropy production, both theoretically and experimentally, by using different methods, such as uncertainty relations \cite{manikandan2021quantitative,das2022inferring,van2020entropy}, waiting-time distributions \cite{skinner2021estimating}, machine learning \cite{otsubo2020estimating}, and stochastic single-trajectory data \cite{roldan2010estimating,otsubo2022estimating,lander2012noninvasive}. While an explosion of research investigate the mean entropy production, there is a lack of  general understanding of the properties of its probability density function (pdf). 
Nevertheless, researchers have obtained the distribution of entropy production for specific settings using analytical~\cite{saha2009entropy,gupta2016fluctuation,gupta2020entropy}, numerical \cite{martynec2020entropy}, and experimental techniques \cite{tietz2006measurement,speck2007distribution,koski2013distribution,manikandan2022nonmonotonic}.

Having an estimate for the moments of the entropy production might be as important as quantifying 
its mean. 
Though a system might be close to equilibrium, it can potentially have large fluctuations of entropy production. In fact, estimating the entire pdf provides information about this variability \cite{ciliberto2017experiments}. From a 
general perspective, our understanding of biological and chemical systems might benefit from the knowledge of fluctuations of any thermodynamic quantity, including entropy production \cite{seara2018entropy,ramaswamy2010mechanics,li2019quantifying,martinez2017colloidal,busiello2021dissipation,dass2021equilibrium}. 
Following this research direction, in \cite{manikandan2020inferring} the authors introduce a method to infer mean and variance of entropy production from short-time experiments, while in \cite{speck2007distribution} these quantities are estimated numerically using differential equations for moments of dissipated heat, following \cite{speck2005integral}. Some studies place bounds on all steady-state currents, including entropy production \cite{dechant2018entropic}, specifically through techniques of linear response theory \cite{pietzonka2016universal}, and large deviation theory \cite{gingrich2016dissipation}. To the best of our knowledge, there exists no theoretical framework to compute the distribution of entropy production which applies to a large class of systems. One of the difficulties encountered is that entropy production is a trajectory-dependent quantity, a property that makes the analytical computation of its statistics beyond the mean a difficult task.

In the study of non-equilibrium systems, modeling of active self-propelled particles has recently been one of the most active fields. These particles break detailed balance via a self-driven term that leads to a wide range of non-equilibrium phenomena resembling various distinctive properties of living systems \cite{marchetti2013hydrodynamics,ramaswamy2017active,prost2015active}. Examples of such phenomena include self assembly \cite{stenhammar2016light,du2019self}, spontaneous segregation \cite{stenhammar2015activity}, and motility induced phase separation \cite{pu2017reentrant,fily2012athermal}. 
The non-equilibrium nature of such systems automatically leads to questions about the properties of their thermodynamic features. Fluctuation theorems in active Ornstein-Uhlenbeck processes \cite{mandal2017entropy,caprini2019entropy,martin2021statistical}, stochastic thermodynamics of active particles \cite{speck2016stochastic,szamel2019stochastic}, their entropy production \cite{chaki2019effects,nardini2017entropy,shankar2018hidden,cao2019effective,skinner2021improved}, heat fluctuations of interacting active particles~\cite{gupta2021heat}, and experimental measurements of uncertainty relations \cite{falasco2016exact} are only a few examples of works performed in this area.

One of the most studied models for active matter components is the run-and-tumble motion. Particles undergoing this dynamics capture the typical homonymous behavior displayed by microorganisms, such as \textit{E. Coli} and \textit{Salmonella}, characterized by driven diffusive dynamics (run) interspersed by random changes of the velocity direction (tumble) \cite{berg2004coli,elgeti2015run}. 
In its simplest form, the model consists of a random walker whose velocity direction is influenced by a dichotomous noise~\cite{demaerel2018active,malakar2018steady}, often referred to as `telegraphic' noise~\cite{rosenau1993random}, and the walker's position is described by the telegrapher's equation (also relevant in electronics \cite{weiss2002some,evans2018run}).
In addition to displaying motion similar to microorganisms, run-and-tumble particles exhibit interesting steady states \cite{dhar2019run,malakar2018steady} which also leads to clustering near the boundaries \cite{elgeti2015run}. Recent studies have also investigated the first passage properties of this system with~\cite{evans2018run} and without stochastic resetting~\cite{angelani2015run}.

Run-and-tumble particles have been shown to have non-zero average entropy production \cite{razin2020entropy}. Due to their popularity, the dynamics might serve as a paradigmatic model to study active matter systems. Here, we start from a discrete-state version of a model for run-and-tumble particles (Sec.~\ref{sec:setup}) and present a graphical method to compute cumulants of the entropy production at any order (Sec.~\ref{sec:entropy-cumulant-method}). Notice that, although the graphical method is employed to study the run-and-tumble setup, it is nevertheless valid for any Markovian system. We proceed to calculate analytically the third cumulant of the entropy production in this model, and verify it through simulations of the system (Sec.~\ref{sec:discrete-run-and-tumble-entropy}). Although obtaining the full distribution remains a lofty goal, our formalism can be, in principle, extended up to a desired precision. 

Furthermore, for the given model, we find an interesting system-size scaling of the entropy production's complementary cumulative distribution function. We emphasize that the approach presented here can be easily generalized to any discrete-state system (both undergoing discrete- and continuous-time evolution) and different boundary conditions. Finally in Sec.~\ref{sec:convergence-continuous}, by implementing a proper coarse-graining procedure, we show that our predictions are compatible with those numerically obtained from the Langevin equation of run-and-tumble particles.

\section{Setup} \label{sec:setup}
\begin{figure}[h]
	\centering
	\includegraphics[width=8.2cm]{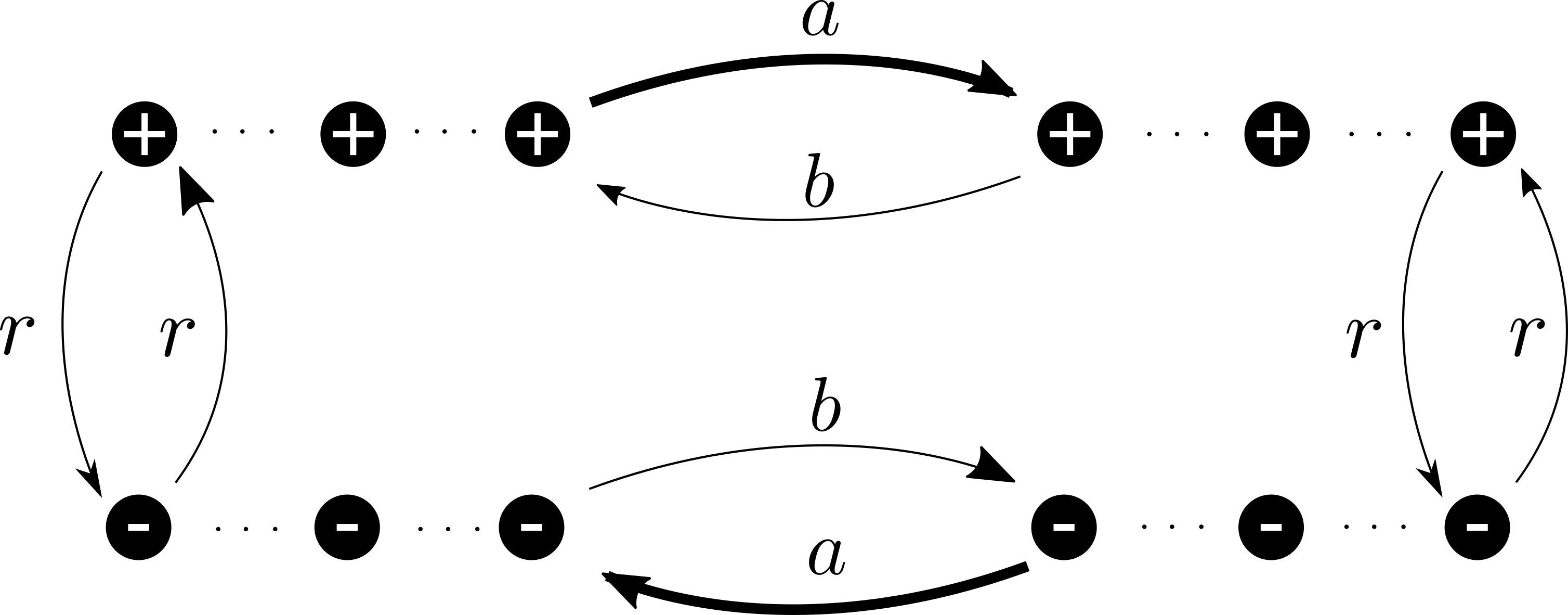}
	\caption{Schematic representation of a run-and-tumble particle. Top layer: + regime. Bottom layer: - regime. The particle hops forward and backward, respectively, with a rate $a$ ($b$) and $b$ ($a$) in the + (-) regime, where $a>b$. Moreover, the particle switches states in between the two layers with a rate $r$.
	} 
	\label{fig:run-and-tumble-representation}
\end{figure}

We consider a run-and-tumble walker on a discrete state-space with reflecting boundary states. The walker hops with a switching rate $r$ between two lanes, representing two different velocity directions in a one-dimensional system. On the upper (lower) lane, the walker hops forward with a rate $a$ ($b$), and backward with a rate $b$ ($a$). Without loss of generality, we consider $a>b$. The schematic diagram describing the system is shown in Fig.~\ref{fig:run-and-tumble-representation}. Thus, the system consists of $2N$ states, of which $N$ states are in the $+$ regime (i.e., upper lane), and the remaining $N$ are in the $-$ regime (i.e., lower lane).  Here `+' and `-' correspond to the direction in which the walker hops on average. \\

The master equation governing the probability of finding a particle in the $i$-th state is \cite{van1992stochastic}
\begin{align}
\dot{P}(i,t) = \sum_{j} W_{ij}~P(j,t),
\label{def:master equation}
\end{align}
where the dot indicates the time-derivative. $W_{ij} \equiv W_{i \leftarrow j}$ is transition rate from the state $j$ to $i$,  $j \neq i$, and the element of the transition rate matrix $\hat{W}$ in the position $(ij)$. Since the probability distribution is normalized at all times, i.e., $\sum_i~P(i,t) = 1$, the sum of 
elements of each column of $\hat{W}$ is zero: $\sum_i W_{ij}=0$ which gives $W_{jj} = - \sum_{i\neq j} W_{ij}$. 
Herein, we consider a system in which if $W_{ij} \neq 0$, so is $W_{ji}$, since there are no unidirectional transitions.

In this paper, we aim to compute fluctuations of entropy production of a run-and-tumble particle. In the case of a discrete state-space, the total entropy production, $\Sigma_{\rm tot}$, at the level of a single trajectory is defined as follows. Given a forward trajectory $\Gamma \equiv \{(i_0,t_0),(i_1,t_1),\dots,(i_{M},t_{M})\}$, where the state $i_k$ is visited at time $t_k$ and changes to state $i_{k+1}$ at time $t_{k+1}$, the asymmetry between the probability of forward and reverse trajectories quantifies the total entropy production 
\cite{seifert2005entropy,seifert2012stochastic,esposito2010entropy,schuster2013nonequilibrium}: 
\begin{align}
\Sigma_{\rm tot}(\Gamma) \equiv \ln  
\dfrac{\mathscr{P}(\Gamma)}{\mathscr{P} (\Gamma^\dagger)},
\end{align}
where $\mathscr{P}(\Gamma)$ and $\mathscr{P} (\Gamma^\dagger)$, respectively, are the probabilities of observing a forward and a time-reversed trajectory. Note that this definition of entropy production is valid when the local detailed balance is satisfied, which we shall assume to be the case \cite{katz1983phase,maes2021local}. In the event of local detailed balance violated, informatic entropy production needs to be considered \cite{li2021steady}.  Following~\cite{busiello2020entropy}, the total entropy production along the trajectory $\Gamma$ reads:
\begin{align}
    \Sigma_{\rm tot}(\Gamma) = \ln\bigg[\frac{P(i_0,t_0)}{P(i_M,t_M)} \prod_{i,j \in \Gamma} \bigg(\frac{W_{ij}}{W_{ji}}\bigg)^{n_{ij}} \bigg],\label{def-tot-ent}
\end{align}
where $P(i_0,t_0)$ and $P(i_M,t_M)$, respectively, are the probabilities of initial and final states of the trajectory, and $n_{ij} \equiv n_{i\leftarrow j} $ counts the number of jumps from the state $j$ to $i$ in the trajectory $\Gamma$. $\Sigma_{\rm tot}$ can be split into two contributions: system entropy production $\Sigma_{\rm sys}$,  and environment entropy production $\Sigma_{\rm env}$. In particular, $\Sigma_{\rm env}$ is associated with the heat dissipated by the particle into the surrounding bath along the trajectory $\Gamma$:
\begin{align}
\Sigma_{\rm env}(\Gamma) \equiv \sum_{i,j \in \Gamma} n_{ij}~\ln
\frac{W_{ij}}{W_{ji}},
\label{def:entropy-environment}
\end{align}
where $j$ precedes $i$ in the trajectory $\Gamma$.
This is the only term that survives in the stationary state, when averaged over many trajectories \cite{busiello2020entropy}. Additionally, for any finite discrete-state systems, $\Sigma_{\rm sys} = \Sigma_{\rm tot} - \Sigma_{\rm env}$ is a boundary term involving the initial and final states for each trajectory, and is the sub-leading contribution to the total entropy production in the long-time limit. 

Since the number of jumps, $n_{ij}$, performed by the run-and-tumble walker is a trajectory-dependent quantity, i.e., it varies from one realization to another, the knowledge of its statistics is required to obtain the fluctuations of the entropy production in Eq.~\eqref{def:entropy-environment}.

\section{Computation of cumulants of entropy production} \label{sec:entropy-cumulant-method}
To compute the various correlations of the number of jumps, we start from a simpler dynamical model: a Markov chain \cite{gardiner1985handbook,busiello2022hyperaccurate}. Unlike the master equation in which time changes continuously, now the time increases in discrete steps, $\Delta t$. In one time increment, the transition probability of the system to jump from the state $j$ to $i$ is $\mathcal{P}(i,t+\Delta t | j,t) \equiv A_{ij} = W_{ij} \Delta t$ for $i\neq j$, and the probability of staying in the state $i$ is $A_{ii} = 1 - \sum_{j \neq i} W_{ji} \Delta t = 1 + W_{ii}\Delta t$. It follows that the sum of the elements of each column is unity, i.e., $\sum_i A_{ij}=1$.
Therefore, the Markov chain equation is
\begin{align}
    P(i,t+\Delta t) = \sum_{j=1}^{2N}~A_{ij}P(j,t)\label{dyn-1}. 
\end{align}
In the limit $\Delta t \to 0$, the above equation~\eqref{dyn-1} reduces to the master equation~\eqref{def:master equation}. Note that herein we are considering time-independent transition rates.

Since the time evolution runs only over times multiple of $\Delta t$, fixing the observation time $T$ is equivalent to fixing the total number of jumps to $T/\Delta t$. In a Markov chain, time is a bookkeeping measure, and therefore, we are able to consider equally spaced time intervals for the trajectory. Hence, the probability of a Markov chain (MC)
trajectory $\Gamma_{\rm MC} \equiv   \{(i_0,t_0),(i_1,t_1),\dots,(i_{M},t_{M})\}$, where $t_{k+1} = t_0 + (k+1) \Delta t$ is:
\begin{align}
    \mathscr{P}(\Gamma_{\rm MC}) \equiv A_{i_M i_{M-1}}A_{i_{M-1} i_{M-2}}\dots~A_{i_{2} i_{1}}A_{i_{1} i_{0}}P(i_0,0),
    \label{def:markov-chain-path-prob}
\end{align}
where $P(i_0,0)$ is the initial probability distribution of the run-and-tumble walker at time $t_0 = 0$.  Note that in Eq.~\eqref{def:markov-chain-path-prob}, it is possible that $i_{k+1} = i_k$ for some $k$'s, i.e., there is a possibility of staying in the same state after the time interval $\Delta t$ which is a consequence of imposing $M$ equally spaced time intervals. When we move back to the master equation, these probabilities of staying in the same state lead to exponential waiting time distributions of times between jumps from one state to another, thereby $\Gamma_{\rm MC}$ converges to $\Gamma$. 

The path probability, $\mathscr{P}(\Gamma_{\rm MC})$, is normalized over all trajectories, i.e., 
\begin{align*}
\sum_{\Gamma_{\rm MC}} \mathscr{P}(\Gamma_{\rm MC}) = \sum_{i_M,i_{M-1},\dots,i_1,i_0} A_{i_M i_{M-1}} A_{i_{M-1} i_{M-2}}\dots~ \nonumber \\
\times~A_{i_{2} i_{1}}A_{i_{1} i_{0}}P(i_0,0) = 1,
\end{align*}
where we used $\sum_i~A_{ij}=1$ for each summation.

The number of jumps performed up to the time $T$ across a link from $\ell$ to $m$ in a trajectory $\Gamma$ is then:
\begin{align}
    n_{m \ell}(\Gamma_{\rm MC}) \equiv \sum_{k=0}^{M-1} \delta_{i_{k+1},m}~\delta_{i_{k},\ell},
\end{align}
where the Kronecker deltas give 1 whenever the system performs jumps from state $\ell$ to $m$.

Let us first compute the average number of jumps over all possible trajectories:
\begin{widetext}
\begin{subequations}
\begin{align}
    \langle n_{m \ell} \rangle_{\Gamma_{\rm MC}} =& \sum_{\Gamma_{\rm MC}} \mathscr{P}(\Gamma_{\rm MC})~ \sum_{k=0}^{M-1} \delta_{i_{k+1},m}~\delta_{i_{k},\ell}\\
    =&\sum_{k=0}^{M-1}\sum_{i_M,i_{M-1},\dots,i_1,i_0}  A_{i_M i_{M-1}}A_{i_{M-1} i_{M-2}}\dots~A_{i_{2} i_{1}}A_{i_{1} i_{0}}P(i_0,0)~\delta_{i_{k+1},m}~\delta_{i_{k},\ell}\label{9beqn}\\
    =&\sum_{k=0}^{M-1} \sum_{i_M,\dots,i_{k+1},i_{k},i_{k-1}} A_{i_M i_{M-1}}\dots A_{i_{k+1} i_{k}}~\delta_{i_{k+1},m}~\delta_{i_{k},\ell}~ A_{i_{k} i_{k-1}}\sum_{i_{k-2},\dots,i_0}A_{i_{k-1} i_{k-2}}\dots~ A_{i_{1} i_{0}}P(i_0,0)\label{9ceqn}\\
    =& \sum_{k=0}^{M-1}\sum_{i_{k-1},\dots,i_0} A_{m \ell}A_{\ell i_{k-1}}A_{i_{k-1} i_{k-2}}\dots A_{i_{1} i_{0}}P(i_0,0)\label{9deqn}\\
    =&\sum_{k=0}^{M-1}A_{m \ell}~P(\ell,k\Delta t)\label{9eeqn}. 
\end{align}
\end{subequations}
\end{widetext}
To go from Eq.~\eqref{9beqn} to \eqref{9ceqn}, we move the Kronecker deltas next to the $\hat{A}$'s matrix elements with the corresponding indices, and identify two groups of indices. Then, the summation over $i_M, \dots, i_{k+2}$ gives $1$ using the property $\sum_i A_{ij}=1$, while the one over the indices $k+1$ and $k$ can be carried out using the Kronecker delta. The resulting expression is in Eq.~\eqref{9deqn}. Finally, we use the Markov chain evolution in Eq.~\eqref{dyn-1} to perform the summation on indices $i_{k-1}$ to $i_0$ to obtain the last equality, Eq.~\eqref{9eeqn}.

Similarly, we compute the correlations between two sets of jumps:
\begin{subequations}
\begin{align*}
	\langle n_{m \ell}~n_{m' \ell'} \rangle_{\Gamma_{\rm MC}} &=   \sum_{\Gamma_{\rm MC}} \mathscr{P}(\Gamma_{\rm MC})\sum_{k=0}^{M-1} \delta_{i_{k+1},m}~\delta_{i_{k},\ell}\nonumber \\
	&\times \sum_{k'=0}^{M-1} \delta_{i_{k'+1},m'}~\delta_{i_{k'},\ell'} \\
	&=\sum_{k=0}^{M-1}\sum_{k'=0}^{M-1}\sum_{i_M,\dots,i_0}~\delta_{i_{k+1},m}~\delta_{i_{k},\ell} ~ \delta_{i_{k'+1},m'}\nonumber \\
	 &\times~\delta_{i_{k'},\ell'}~ A_{i_M i_{M-1}}\dots~A_{i_{1} i_{0}} P(i_0,0). 
\end{align*}
\end{subequations}
We split the second summation over $k'$ depending on three different scenarios: 1) $k'<k$, 2) $k'=k$, and 3) $k'>k$.  Performing similar calculations as in the case of the first moment, we obtain, for $k'<k$,
\begin{align}
\langle n_{m \ell}~n_{m' \ell'} \rangle_{\Gamma_{\rm MC}} =\sum_{k=0}^{M-1}\sum_{k'=0}^{k-1}&A_{m \ell}~\mathcal{P}(\ell,k\Delta t|m',(k'+1)\Delta t)\nonumber \\
 \times &A_{m' \ell'}~P(\ell',k'\Delta t),\label{klkp}
\end{align} 
for $k<k'$,
\begin{align}
   \langle n_{m \ell}~n_{m' \ell'} \rangle_{\Gamma_{\rm MC}} =\sum_{k=0}^{M-1}\sum_{k'=k+1}^{M-1}&A_{m' \ell'}~\mathcal{P}(\ell',k'\Delta t|m,(k+1)\Delta t)\nonumber \\
   \times&A_{m \ell}~P(\ell,k\Delta t),\label{kgkp}
\end{align}
and for $k'=k$,
\begin{align}
     \langle n_{m \ell}~n_{m' \ell'} \rangle_{\Gamma_{\rm MC}} &=\sum_{k=0}^{M-1}A_{m \ell}~P(\ell,k\Delta t) ~\delta_{m,m'}\delta_{\ell,\ell'}.\label{kekp}
\end{align}
Combining the above three contributions, Eqs.~\eqref{klkp}, \eqref{kgkp}, and \eqref{kekp}, finally we obtain: 
\begin{align}
   &\langle n_{m \ell}~n_{m' \ell'} \rangle_{\Gamma_{\rm MC}} = \nonumber \\
   &\sum_{k=0}^{M-1}\bigg[\sum_{k'=0}^{k-1}A_{m \ell}~\mathcal{P}(\ell,k\Delta t|m',(k'+1)\Delta t) A_{m' \ell'}~P(\ell',k'\Delta t)&\nonumber\\ &+\sum_{k'=k+1}^{M-1}A_{m' \ell'}~\mathcal{P}(\ell',k'\Delta t|m,(k+1)\Delta t)A_{m \ell}~P(\ell,k\Delta t)\nonumber\\&+A_{m \ell}~P(\ell,k\Delta t) \delta_{m,m'}\delta_{\ell,\ell'}\bigg].
   \label{second-moment-terms}
\end{align}
Such calculations become tedious on proceeding to higher order correlations. However, we present a graphical method to scale up the calculations to any order of correlations of the number of jumps. For a given correlation, we first determine the set of all possible time-orderings of $k$-s, i.e. the times at which a specific jump takes place. For the first moment, there is only one jump considered, hence no ordering is required. For the correlations between two sets of jumps, say $\{m,\ell\}$ and $\{m',\ell'\}$, happening at times $k\Delta t$ and $k'\Delta t$ respectively, as mentioned earlier, the possible permutations are $k<k'$, $k>k'$, and $k=k'$. Once all the orderings are listed, the set of states are graphically located according to the orderings. For example, corresponding to $k<k'$, the set of states $\{m,\ell\}$ appears before in time than the set of states $\{m',\ell'\}$. Notice that in what follows, we consider the time-axis from right to left to be consistent with the ordering at which propagators appear. 

\begin{figure}[t]
	\centering
	\includegraphics[width=8.2cm]{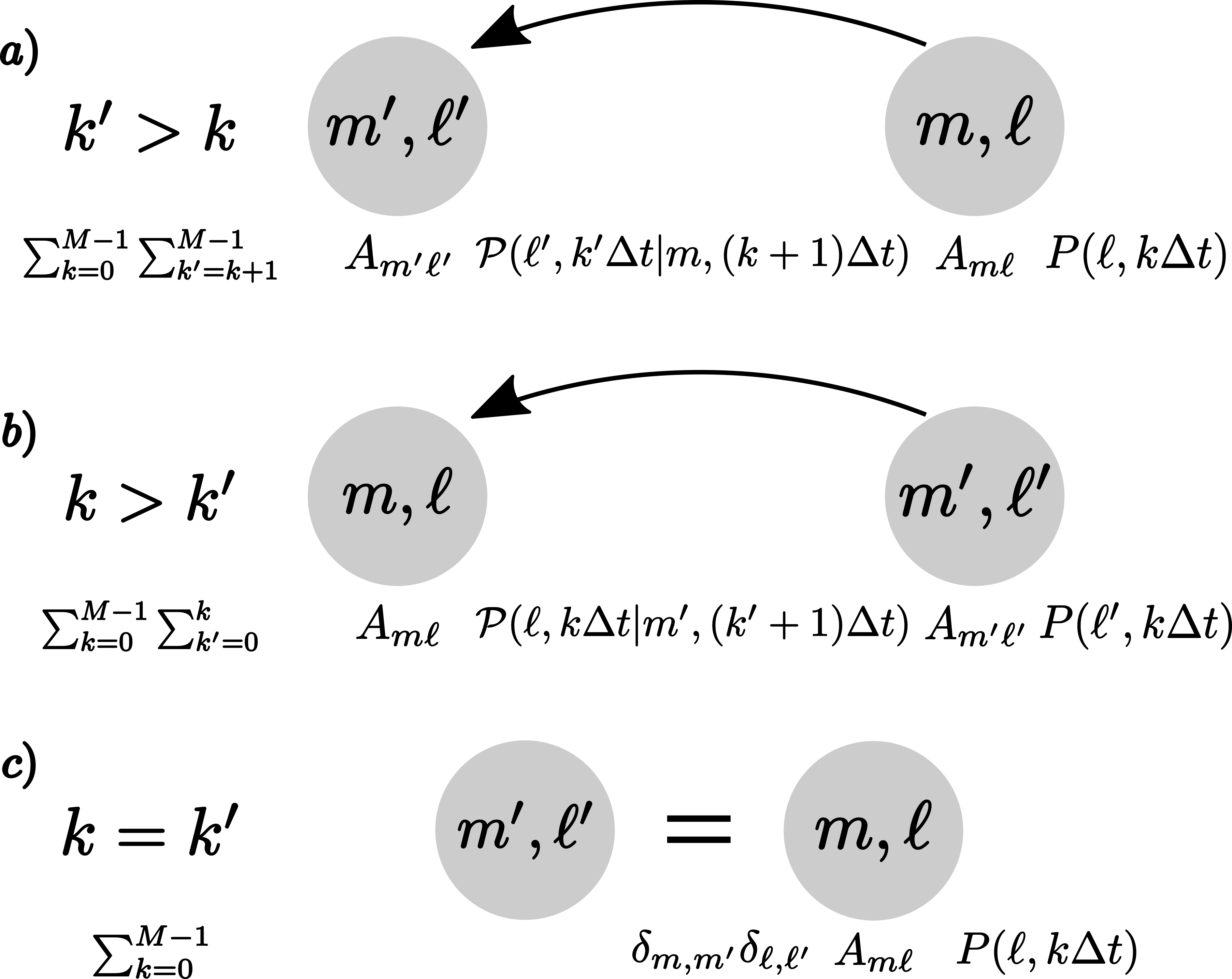}
	\caption{Graphical representation for the computation of second order correlation for number of jumps. a) $k'>k$, b) $k>k'$, and c) $k'=k$. Circle indicates the set of states corresponding to the summation label, either $k$ or $k'$. The arrow and equality, respectively, correspond to the transition probability from left set of states to the right ones, and the Kronecker deltas equating the set of states.  }
	\label{fig:graphical-method-second-moment}
\end{figure}

Fig.~\ref{fig:graphical-method-second-moment} shows 
possible orderings for the second order correlation.

For $k'>k$ (see Fig.~\ref{fig:graphical-method-second-moment}a), 
the rightmost circle carries a contribution from its starting state, $\{m,l\}$, at time $k\Delta t$. The contribution is equal to its probability $A_{m \ell} P(\ell, k\Delta t)$. Then the system moves towards the left circle, which is associated to the final set of states in this scenario. This transition comes with its propagator: $A_{m' \ell'} \mathcal P(\ell', k' \Delta t|m, (k+1) \Delta t)$.
Finally the summation runs over all possible indices $k$ and $k'$ with the prescribed ordering ($k'>k$ in this case). Hence, we can immediately write the contribution to the second order correlation as given by Eq.~\eqref{kgkp}. Similarly, we can write the contributions for $k'=k$, and $k'<k$.

\begin{figure*}[t]
	\centering
	\includegraphics[width=2\columnwidth]{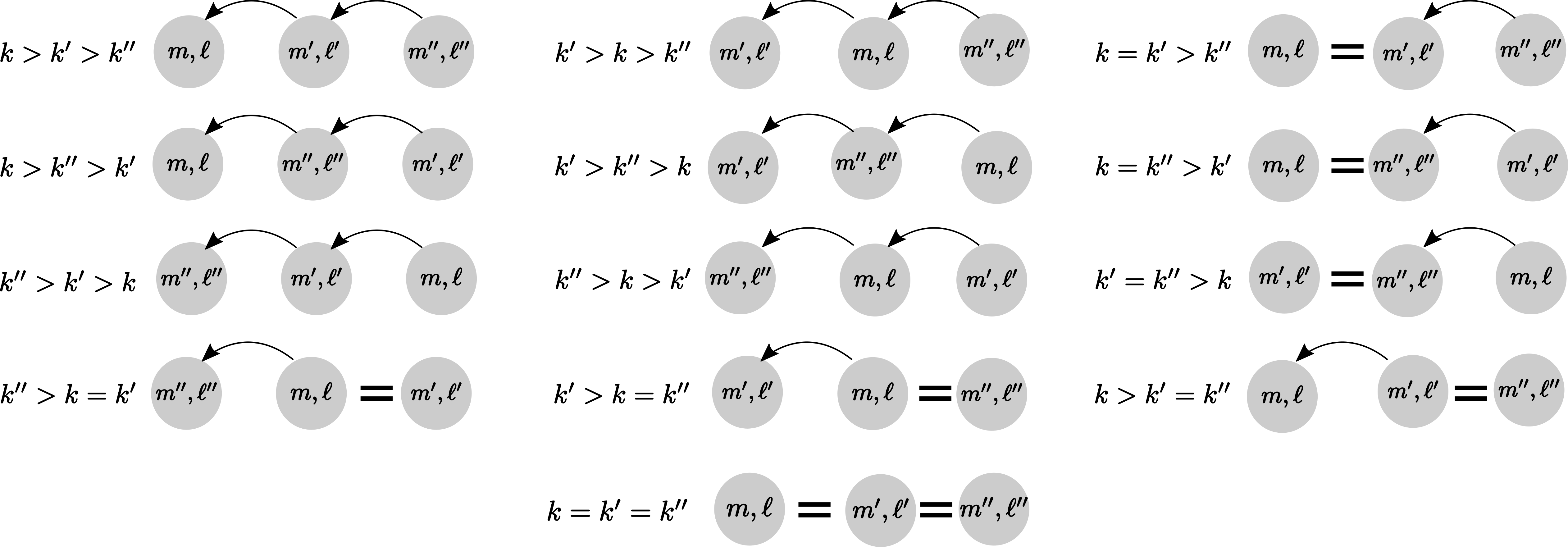}
	\caption{Possible orderings in the graphical method for the third order correlation of number of jumps.}
	\label{fig:graphical-method-third-moment}
\end{figure*}

For the third order correlation, repeating the graphical procedure leads to 13 possible orderings with three $k$-s indices (i.e., $k,k',k''$). We show all the orderings in Fig.~\ref{fig:graphical-method-third-moment}. Writing down the summation terms according to the graphical rules, we find them to be equal to those obtained from the full calculation
In order to avoid clutter, we relegate the detailed form of the third order jump correlation to Appendix~\ref{appendix:third-jump-correlation}.

To move back from a Markov chain to a master equation description, we rewrite the transition probability as $A_{m \ell} = W_{m \ell} \Delta t$, $m \neq l$, and take the limite $\Delta t \to 0$. Thus, each summation over $k$-s appearing in the jump correlations is converted into an integral over time, $t$. Although the calculations shown above are valid for an arbitrary initial condition, in what follows, we focus on the case in which the system starts from an initial steady state distribution, $P(i_0,t_0) = P^{\rm st}(i_0)$ and $P(i_M,T) = P^{\rm st}(i_M)$. Thus, the jump correlations in Eqs.~\eqref{9eeqn} and \eqref{second-moment-terms} in the continuous-time limit read:
\begin{align}
\langle n_{m \ell} \rangle_\Gamma =& \int_0^{T} {\rm d}t~W_{m \ell}~P^{\rm st}(\ell), \label{def:first-jump-moment} \\
\langle n_{m \ell}n_{m' \ell'} \rangle_\Gamma =& \int_{0}^{T}{\rm d}t~ \bigg(\int_{0}^{t}{\rm d}t'~W_{m \ell}~\mathcal P(\ell,t|m',t')~W_{m' \ell'}  \nonumber \\
&\times P^{\rm st}(\ell') + \int_{t}^{T} {\rm d}t'~W_{m' \ell'}~\mathcal P(\ell',t'|m,t)\nonumber \\
&\times W_{m \ell}~P^{\rm st}(\ell) \bigg) \nonumber \\
&+ \int_{0}^{T}{\rm d}t~W_{m \ell}~P^{\rm st}(\ell)~\delta_{\ell,\ell'}~\delta_{m,m'}, \label{def:second-jump-moment}
\end{align}
where $\mathcal{P}(\ell',t'|m,t)$ is the probability to be in the state $\ell'$ at time $t'$, starting from the state $m$ at time $t$, computed from the master equation. The same limit can be computed for the third moment, as shown in Appendix~\ref{appendix:third-jump-correlation}.

The integration on the right-hand side of first jump moment, Eq.~\eqref{def:first-jump-moment}, yields:
\begin{align}
\langle n_{m \ell} \rangle_\Gamma=T~W_{m \ell}~P^{\rm st}(\ell), \label{eval-fm}
\end{align}
whereas the computation of higher order jump moments requires the knowledge of the transition probability: $\mathcal P(i',t'|i,t)$. To this end, we use the eigenvector expansion of the transition rate matrix $\hat{W}$ to compute such quantity. The master equation can be written in a compact matrix form:
\begin{align}
|\dot{P}(t)\rangle &= \hat{W}~|P(t)\rangle,
\label{P-t}
\end{align}
where $|{P}(t)\rangle = [P(1,t),P(2,t),\dots]^\top$ is the probability vector, and $\top$ is the matrix transpose operator. The solution of the above linear differential equation~\eqref{P-t}, given an initial state vector $|P(t_0)\rangle$, is

\begin{align}
|P(t)\rangle &= e^{\hat W(t-t_0)}|P(t_0)\rangle. \label{p-eqn-vec}
\end{align}

Let $\langle \psi_\mathfrak{j}|$ and $|\phi_\mathfrak{j}\rangle$, respectively, be the $\mathfrak{j}$-th left and right eigenvectors of the transition rate matrix, $\hat{W}$, corresponding to eigenvalue $\lambda_\mathfrak{j}$. The left and right eigenvectors satisfy the normalization condition \cite{van1992stochastic}:
\begin{align}
\langle \psi_{\mathfrak{j}}|\phi_{\mathfrak{j}'}\rangle  = \delta_{\mathfrak{j},\mathfrak{j}'}.
\end{align}
Expanding the right-hand side of Eq.~\eqref{p-eqn-vec} in the eigenbasis of $\hat{W}$ gives:
\begin{align}
|P(t)\rangle &= \sum_{\mathfrak{j}}\langle \psi_\mathfrak{j}|P(t_0)\rangle~e^{-\lambda_\mathfrak{j}(t-t_0)}~|\phi_\mathfrak{j}\rangle,
\end{align}
where $0=\lambda_1 < \Re(\lambda_2)\leq \Re(\lambda_3)\leq\dots \leq \Re(\lambda_{2N})$, where $\Re(\lambda_\mathfrak{j})$ represents the real part of $\lambda_\mathfrak{j}$.
The system considered here reaches a steady-state in the long-time limit, $|P(t\to \infty)\rangle \to |P^{\rm st}\rangle$ which is the right eigenvector corresponding to $\lambda_\mathfrak{j=1}=0$ eigenvalue, i.e., $|\phi_\mathfrak{j=1}\rangle$. Since $\langle \psi_{\mathfrak{j=1}}|$ is a row vector with all entries equal to 1, it gives the condition $\langle \psi_\mathfrak{j=1}|P(t_0)\rangle=\sum_i P(i,t_0)=1$.

In particular, if the initial state vector $|P(t_0)\rangle$ is a column vector of all zeros except 1 at $i_0$-th location, then the system is in the state $i_0$ at time $t_0$. Let us call this vector $|i_0\rangle$. Then, the probability of the system to be in state $i$ at time $t$ given the initial state $i_0$ at time $t_0$, $\mathcal{P}(i,t|i_0,t_0) \equiv \langle i|P(t)\rangle$, can be written as:    
\begin{align}
\mathcal{P}(i,t|i_0,t_0) &= \sum_{\mathfrak{j}}c_{\mathfrak{j}}(i_0)~e^{-\lambda_\mathfrak{j}(t-t_0)}~\phi_\mathfrak{j}(i), \label{exp-P}
\end{align}
where we defined the projection of the left eigenvector onto the initial state as the coefficients of the expansion, i.e., $c_{\mathfrak{j}}(i_0)\equiv \langle \psi_\mathfrak{j}|i_0\rangle$. Similarly, we define $\phi_\mathfrak{j}(i)\equiv \langle i|\phi_\mathfrak{j}\rangle$.

Using the eigenvector expansion, Eq.~\eqref{exp-P}, in the integrals appearing in Eq.~\eqref{def:second-jump-moment}, we obtain
\begin{widetext}
\begin{align}
\langle n_{m \ell} n_{m' \ell'} \rangle_\Gamma &=  W_{m\ell}~W_{m'\ell'}~\bigg[P^{\rm st}(\ell)~P^{\rm st}(\ell') T^2 + \sum_{\mathfrak{j}>1} [ c_\mathfrak{j}(m)~\phi_\mathfrak{j}(\ell')~P^{\rm st}(\ell) + c_\mathfrak{j}(m')~\phi_\mathfrak{j}(\ell)~P^{\rm st}(\ell')]\nonumber \\
&\times \frac{1}{\lambda_\mathfrak{j}}\bigg(T - \frac{1}{\lambda_\mathfrak{j}}(1-e^{-\lambda_\mathfrak{j} T})\bigg)\bigg] + \delta_{\ell,\ell'}~\delta_{m,m'}~W_{m \ell}~P^{\rm st}(\ell)~T.
\label{eqn:second_jump_moment_final_formula}
\end{align}
\end{widetext}

\begin{figure*}
	\centering
	\includegraphics[width=18cm]{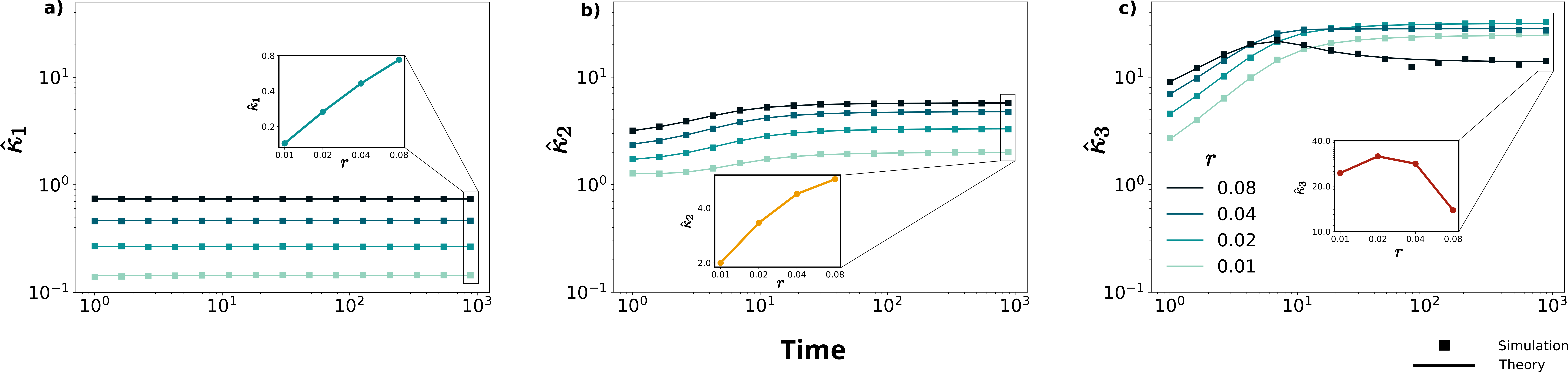}
	\caption{Scaled cumulants of entropy production. Dots: Numerical simulation. Lines: theoretical predictions. Number of states in each regime $N=8$, with transition rates $a=1.0$, and $b=0.1$. Inset shows the variation of $\hat \kappa_{1,2.3}$ with different switching rates $r$. Here the averaging is performed over $10^5$ trajectories (generated using the Gillespie algorithm). In each panel, the color intensity increases with $r$.}
	\label{fig:r-scaling}
\end{figure*}

For the third order jump correlation, let us consider an example of one of the orderings,  $k<k'<k''$, 
with $\{m,\ell\}$, $\{m',\ell'\}$, and $\{m'',\ell''\}$ being the set of states corresponding to $k$, $k'$, and $k''$ respectively. The contribution of this ordering is:
\begin{align}
\langle n_{m,\ell}n_{m',\ell'}n_{m'',\ell''}\rangle_\Gamma =&W_{m'' \ell''}~W_{m' \ell'}~W_{m \ell}~P^{\rm st}(\ell)\nonumber \\ 
&\times \sum_{\mathfrak{j}_1,\mathfrak{j}_2} \big[\phi_{\mathfrak{j}_1}(\ell'')~c_{\mathfrak{j}_1}(m')\nonumber \\
&\times \phi_{\mathfrak{j}_2}(\ell')~c_{\mathfrak{j}_2}(m)~\mathcal{T}_{\mathfrak{j}_1,\mathfrak{j}_2}\big],
\end{align}
where $\mathcal{T}_{\mathfrak{j}_1, \mathfrak{j}_2}$
represents the solution to the integral over time appearing in the third order jump correlation (see Appendix \ref{appendix:third-jump-correlation}). It is given by:
\begin{align}
\mathcal{T}_{\mathfrak{j}_1,\mathfrak{j}_2} \equiv \frac{\lambda_{\mathfrak{j}_2}^2 \left(1-T \lambda_{\mathfrak{j}_1}-e^{-T \lambda_{\mathfrak{j}_1}}\right)-\lambda_{\mathfrak{j}_1}^2 \left(1-T \lambda_{\mathfrak{j}_2}-e^{-T \lambda_{\mathfrak{j}_2}}\right)}{\lambda_{\mathfrak{j}_2}^2 \left(\lambda_{\mathfrak{j}_1}-\lambda_{\mathfrak{j}_2}\right) \lambda_{\mathfrak{j}_1}^2}. \label{third-moment-time-integral}
\end{align}
When $\lambda_{\mathfrak{j1}} = \lambda_{\mathfrak{j2}}$, Eq. (\ref{third-moment-time-integral}) is indeterminate. Taking L'H\^{o}pital's rule, we find 
\begin{align}
    \lim_{\lambda_{\mathfrak j_1} \to \lambda_{\mathfrak j_2} } \mathcal{T}_{\mathfrak j_1,\mathfrak j_2} =\dfrac{T \lambda _{\mathfrak j_1}+e^{-T \lambda _{\mathfrak j_1}} \left(T \lambda _{\mathfrak j_1}+2\right)-2}{\lambda _{\mathfrak j_1}^3}.
\end{align}
Eq \eqref{third-moment-time-integral}  is also indeterminate when either of the eigenvalues is zero. In such circumstances, applying L'H\^{o}pital's rule twice, we obtain its limiting value. 
As an example, the limit $\lambda_{\mathfrak{j}_2} \to 0$ with $ \lambda_{\mathfrak{j}_1} \neq 0$ results in the integral having the form
\begin{align}
    \lim_{\lambda_{\mathfrak{j}_2} \to 0} \mathcal{T}_{\mathfrak j_1,\mathfrak j_2} = \dfrac{2 \left(1-e^{-T \lambda _{\mathfrak j_1}}\right)+T \lambda _{\mathfrak j_1} \left(T \lambda _{\mathfrak j_1}-2\right)}{2 \lambda _{\mathfrak j_1}^3}.
    \label{lambda1 zero integral}
\end{align}
The solution \eqref{lambda1 zero integral} is similar for $\lambda_{\mathfrak j_2}$ if $\lambda_{\mathfrak{j}_1} \to 0$ with $\lambda_{\mathfrak{j}_2} \neq 0$. If both eigenvalues are zero, i.e., $\lambda_{\mathfrak j_1} = \lambda_{\mathfrak j_2} = 0$, Eq.~\eqref{third-moment-time-integral} results in
\begin{align}
    \lim_{\lambda_{\mathfrak j_1}, \lambda_{\mathfrak j_2} \to 0} \mathcal{T}_{\mathfrak j_1,\mathfrak j_2} = \frac{T^3}{6}.
\end{align}

Notice that for all the terms in which two events happen at the same time, for example, $k_1 = k_2<k_3$, the contribution to the $n$-th order jump correlation can be written in terms of the $n-1$-th order one (see Appendix \ref{appendix:third-jump-correlation}). Iterating through all possible orderings and using the solution of the time integral in Eq.~\eqref{third-moment-time-integral}, we can obtain the complete third order correlation for the number of jumps.

The moments and the cumulants of the environmental entropy production can be calculated from the corresponding moments and correlations for number of jumps. Indeed, for instance,
\begin{align}
\kappa_1(T) \equiv \langle \Sigma_{\rm env}(T) \rangle = \sum_{i,j} \langle n_{ij} \rangle_\Gamma~ \ln \frac{W_{ij}}{W_{ji}},
\end{align} 
where the $i,j$ indices run over all $2N$ states.
Scaled cumulants can then be defined as
\begin{subequations}
\label{def:kappas}
\begin{align}
    &\hat \kappa_1(T)  \equiv \frac{\kappa_1(T)}{T}, \label{def:first-kappas}\\
    &\hat \kappa_2(T) \equiv \frac{\kappa_2(T)}{T} \equiv \frac{1}{T}\big(\langle \Sigma_{\rm env}^2 \rangle - \langle \Sigma_{\rm env} \rangle^2 \big), \label{def:second-kappa} \\
    &\hat \kappa_3(T) \equiv \frac{\kappa_3(T)}{T} \equiv \frac{1}{T}\big( \langle \Sigma_{\rm env}^3 \rangle - 3\langle \Sigma_{\rm env}^2 \rangle\langle \Sigma_{\rm env} \rangle + 2\langle \Sigma_{\rm env} \rangle^3 \big),
\end{align}
\end{subequations}
where the time dependence of $\Sigma_{\rm env}$ has been omitted for convenience.

\begin{figure*}[ht]
	\centering
	\includegraphics[width=14cm]{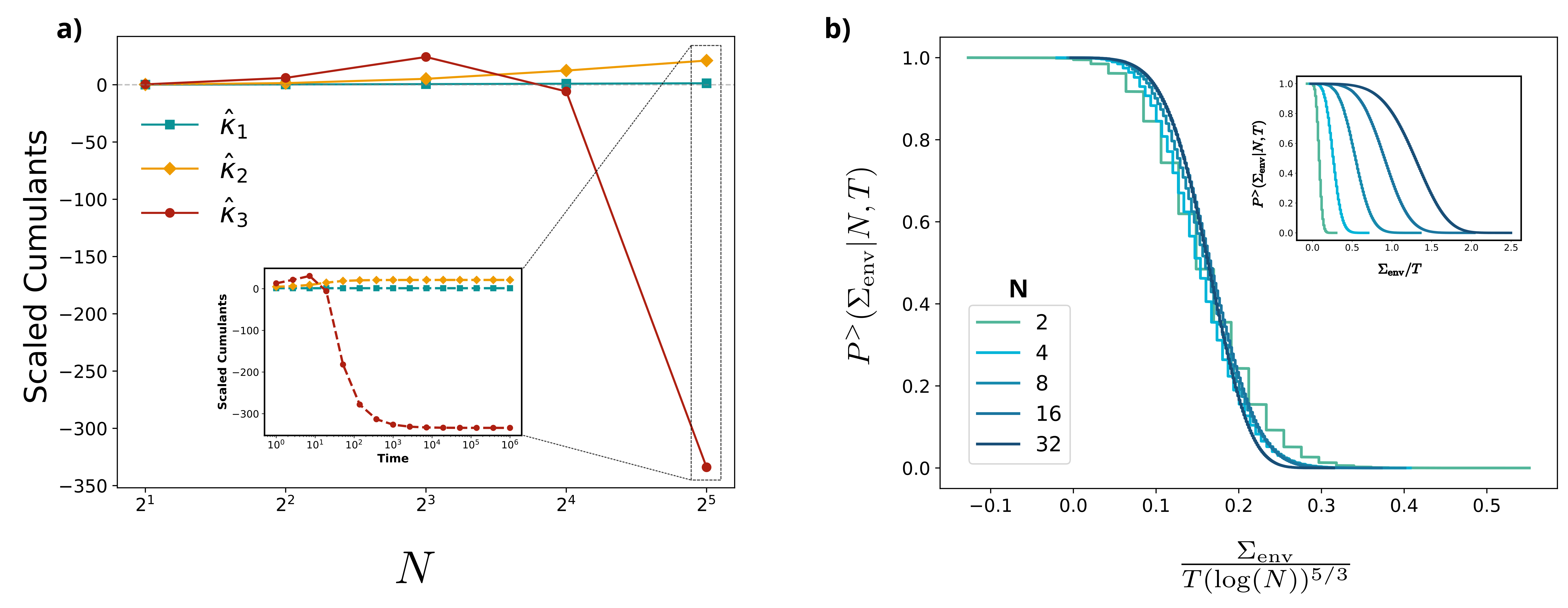}
	\caption{a) Scaling of cumulants ($\hat \kappa_{1,2,3}$) of entropy production for different number of states of the discrete run-and-tumble model. Points are obtained using analytical expressions (lines serve as a visual aid to connect the dots). Inset shows the evolution of scaled cumulants against time for the system with $N = 2^5$. b) Collapse of the complementary cumulative density function (c-cdf) of entropy production, $P^{>}(\Sigma_{\rm env}|N,T)$, for different number of nodes in the discrete run-and-tumble model. Entropy production at time $T = 200$ is obtained from numerical simulation using $10^6$ trajectories initialized from stationary state. Inset shows the uncollapsed c-cdf, without appropriate rescaling with number of nodes. Parameters for both panels are $a=1.0$, $b=0.1$ and $r=0.05$.}
	\label{fig:n-scaling}
\end{figure*}

We can immediately see that the first jump moment scales linearly with time as seen from Eq.~(\ref{eval-fm}), so does the average entropy production, $\langle \Sigma_{\rm env}(T) \rangle$. Concerning the second cumulant, the first term on the right-hand side of Eq.~\eqref{eqn:second_jump_moment_final_formula} scales with $T^2$. However, this term cancels out when evaluating the cumulant, since it is equal to $\langle n_{m \ell} \rangle \langle n_{m' \ell'} \rangle$ [see Eq.~\eqref{eval-fm}]. Hence, in the long-time limit, i.e., $T \gg \max(1/\lambda_{\mathfrak{j}}$, $1<\mathfrak j \leq 2N )$ ($\lambda_1=0$ corresponding
to the stationary state), the second and third terms on the right-hand side of Eq.~\eqref{eqn:second_jump_moment_final_formula} grow linearly with the observation time $T$. Therefore, in this limit, the second cumulant defined in Eq.~(\ref{def:second-kappa}) becomes
\begin{align}
    \langle \Sigma_{\rm env}^2 \rangle - \langle \Sigma_{\rm env} \rangle^2 \approx 
    T \times \sum_{i,j,k,l} f(i,j,k,l)
    \label{variance-scaling-general}
\end{align}
where $f(i,j,k,l)$ is a function that depends only on the states of the system but not on time, and can be readily determined from Eq.~\eqref{eqn:second_jump_moment_final_formula}. Hence, Eqs.~\eqref{def:first-kappas} and \eqref{def:second-kappa} give that, in the long-time limit,
\begin{align}
    &\hat{\kappa}_1 = {\rm constant}, \\
    & \hat{\kappa}_2 = {\rm constant}.
\end{align}
Therefore, in any finite discrete system with bidirectional time-independent transition rates, at large times, the mean and the variance of the environmental entropy production scale linearly with time. This agrees with previous results by Lebowitz and Spohn \cite{lebowitz1999gallavotti} and hence, we expect all cumulants to scale linearly with time at large times, for both discrete and continuous state systems.

\section{Entropy Production in the run-and-tumble model}
\label{sec:discrete-run-and-tumble-entropy}
In the proposed framework, we return to the run-and-tumble model shown in Fig.~\ref{fig:run-and-tumble-representation}. The transition rate matrix $\hat W$ has the following elements: $W_{i_+, i_+ +1} = b,~W_{i_+,i_+-1} = a,~W_{i_-,i_-+1} = a,~W_{i_-,i_- - 1} = b,~  W_{i_+,i_-} = W_{i_-,i_+} = r$ with zero cross transition rates between two layers, and $W_{1_{\pm},0_{\pm}} = 0$ and $W_{N_{\pm},N_{\pm} + 1} = 0$, where the subscript $\pm$ again denotes the respective regime of the states.

We analytically calculate the cumulants of the environmental entropy production for this system in the stationary state (up to the third cumulant, for the sake of simplicity). Furthermore, we simulate the dynamics, generating trajectories that start from the steady state, and numerically compute the entropy production. Figure~\ref{fig:r-scaling} shows a comparison of the scaled cumulants of $\Sigma_{\rm env}$ for various values of switching rate $r$, obtained from analytical results with their numerical simulation counterpart. We find that each scaled cumulant reaches a stationary value in the long-time limit. The non-vanishing value of the third cumulant reflects the fact that the probability density function of the entropy production is asymmetric about its mean value.

Figure~\ref{fig:r-scaling} also shows how $\hat\kappa_{1,2,3}$ change with increasing $r$ as a function of time. We notice that the scaled average entropy production increases when the switching rate increases. This effect can be understood by realizing that, when $r \to 0$, each layer will relax to an equilibrium distribution, hence generating no entropy into the environment on average. Hence, when $r$ increases, the system starts to feel the non-equilibrium condition that is generated by the presence of two regimes, $+$ and $-$, and the entropy production increases. Due to the same reason, the variance of the entropy production also increases with increasing $r$. As a second observation, the third cumulant is consistently far from zero, stressing the non-Gaussianity of the pdf of entropy production.

It is also important to analyze the scaling of the cumulants with the number of nodes, in order to investigate how the distribution of entropy production changes as a function of the system size. 
We analytically compute the scaled cumulants $\hat{\kappa}_{1,2,3}$ for various system sizes, starting from the steady state, and show them in Fig.~\ref{fig:n-scaling}a, while the inset shows scaled cumulants against time for the system with $N$ = 32.
We observe that the first two cumulants increase with $N$. Conversely, the sign of the third cumulant changes with $N$, due to the shift of mode of the distribution of entropy production with respect to its mean, i.e., skewness is positive when the mean is greater than the mode, and negative otherwise.

Then, we numerically simulate the system for different $N$, starting from the steady state, to compute the complementary cumulative density function (c-cdf), $P^{>}(\Sigma_{\rm env} | N,T)$, of $\Sigma_{\rm env}$, defined as
\begin{align}
    P^{>}(\Sigma_{\rm env} | N,T) \equiv \int_{\Sigma_{\rm env}}^{\infty}ds~ p_{\Sigma}(s | N,T). \label{ccdf-def}
\end{align}
Here, $p_{\Sigma}(s | N,T)$ is the probability density function of finding the environmental entropy production to be equal to $s$ at time $T$ for a system of $N$ sites in each regime. We find the leading order scaling with $N$ of the c-cdf of $\Sigma_{\rm env}$ to be $(\ln\left(N\right))^{5/3}$. We note that this is a preliminary observation about the distribution of entropy production and warrants further studies to understand its origin.

In Fig.~\ref{fig:n-scaling}b, we show the collapse of different c-cdf for an increasing number of nodes, $N$. Clearly, there are also sub-leading contributions to the scaling of the moments that play a role in determining the behavior of the third cumulant shown in Fig.~\ref{fig:n-scaling}a.

Notice that increasing the number of states without scaling the rates by $N$ does not correspond to the correct continuum limit~\cite{busiello2019entropy}.
In the next section, we present how to generalize our findings to the case of a run-and-tumble particle in a continuous domain by considering appropriate rescaling of the transition rates.

\section{Convergence to continuous run-and-tumble model} \label{sec:convergence-continuous}

Let us start from the description of a particle experiencing run-and-tumble dynamics in a $1D$ continuous space $[-L,L]$, with reflecting boundary conditions. The Langevin equation describing this dynamics is \cite{malakar2018steady}:
\begin{align}
    \dot{x} = v~\sigma(t) + \sqrt{2D}\eta(t), \label{lang-eqn}
\end{align}
where $\sigma(t) = \pm 1$ is a dichotomous noise that switches between $+1$ and $-1$ with a constant rate $r$, $v$ the bare velocity of the particle in either direction in the absence of thermal noise, $D\equiv{k_{\rm B}\mathbb T/\gamma}$ the diffusion constant (for $k_{\rm B}$ the Boltzmann's constant, $\mathbb T$ the temperature, and $\gamma$ the dissipation constant), and $\eta(t)$ is the Gaussian white noise with zero mean and unit variance.
The corresponding Fokker-Planck equation reads:
\begin{subequations}
\label{rt-fpe}
\begin{align}
    &\partial_t \rho_+ = -v~\partial_x \rho_+ + D \partial^2_x \rho_+ - r (\rho_+ - \rho_-),  \\
    &\partial_t \rho_- = +v~\partial_x \rho_- + D \partial^2_x \rho_- - r(\rho_- - \rho_+),
\end{align}
\end{subequations}
where $\rho_+$ and $\rho_-$, respectively, are the probability density functions for the system to be in the state $\sigma = +1$ and $\sigma = -1$, respectively, at the position $x$ and time $t$ \cite{malakar2018steady}. For convenience, we have omitted the position and time dependence from $\rho_{\pm}(x,t)$.

Let us now go back to our original discrete-state description. The particle can move either in the upper or in the lower $1D$ lattices, i.e., lanes, with a rate of switching between the lanes equal to $r$. The master equation associated solely with the motion along the upper lane ($+$), ignoring the switching between lanes, is:
\begin{equation}
    \dot{P}(i_+,t) = a P(i_+-1,t) + b P(i_++1,t) - (a+b) P(i_+,t).
    \label{eqUPPER}
\end{equation}
A similar equation holds also for the lower lane, interchanging $a$ with $b$, following the model sketched in Fig.~\ref{fig:run-and-tumble-representation}). In order to map this dynamics to a continuous space, we introduce the information that the system exists in a $1D$ box, $[-L,L]$. Hence, as we increase the number of states $N$ in each lane, the spacing between the states has to decrease. In particular, let the spacing between the states $\delta \equiv 2L/N$. Considering again the upper lane, employing this mapping, the spatial position of the particle, $x = i_+ \delta$, and the probability density function transforms as follows: $\rho_+(x) = P(i_+)/\delta = P(x/\delta)/\delta$.

A standard Kramers-Moyal expansion \cite{gardiner1985handbook} on Eq.~\eqref{eqUPPER}, taking $\delta$ as small parameter in the limit $N \to +\infty$, up to the second order, gives:
\begin{equation}
    \partial_t \rho_+ = -(a-b)\delta\frac{\partial \rho_+}{\partial x} +  \frac{(a+b)}{2}\delta^2 \frac{\partial^2 \rho_+}{\partial x^2}.
\end{equation}
Performing the same expansion on the lower lane dynamics as well, and adding the switching process between these two regimes, we can compare the resulting set of coupled differential equation with Eq.~\eqref{rt-fpe}. The matching between these two dynamical evolution becomes exact in the $N \to +\infty$ limit, when the following scaling holds:
\begin{subequations}
\begin{align}
a &= \frac{N}{4L}\bigg(\frac{DN}{L} + v \bigg), \\
b &= \frac{N}{4L}\bigg(\frac{DN}{L} - v \bigg).
\end{align}
\label{discrete-continuous-scaling-relation}
\end{subequations}
It is indeed always true that when performing the continuum limit starting from a discrete-state process, the rates have to properly scaled with the number of states.

Let us now compute the thermodynamics of the continuous process. Given the Langevin equation~\eqref{lang-eqn}, the amount of the heat absorbed by the run-and-tumble particle from the heat bath during an observation time, $T$, is~\cite{sekimoto2010stochastic}:
\begin{align}
 Q \equiv \int_0^T~{\rm d}\tau~\big[\sqrt{2D\gamma^2}\eta(\tau) - \gamma \dot{x}(\tau)\big] \circ \dot{x}(\tau),
\end{align}
where $\circ$ denotes the Stratonovich product.
In the absence of switching, the system satisfies detailed balance and reaches equilibrium. In the presence of switching, effectively, there is only an additional stochastic force on the system with respect to the equilibrium scenario. Hence, we expect the local detailed balance to hold. Thus, the environmental entropy production is~\cite{seifert2005entropy}:
\begin{align}
    S_{\rm env}(T) &= \frac{\gamma}{\mathbb T} \int_0^T~{\rm d}\tau~\big[\dot{x}(\tau) - \sqrt{2D}\eta(\tau)\big]\circ \dot{x}(\tau).
    \label{def-entropy-continuous}
\end{align}
where we use the Einstein relation, $D = k_B \mathbb T/\gamma$ for $k_B = 1$.
This system is known to have non zero mean entropy production rate~\cite{razin2020entropy}. 

\begin{figure}[h]
    \centering
    \includegraphics[width=8.2cm]{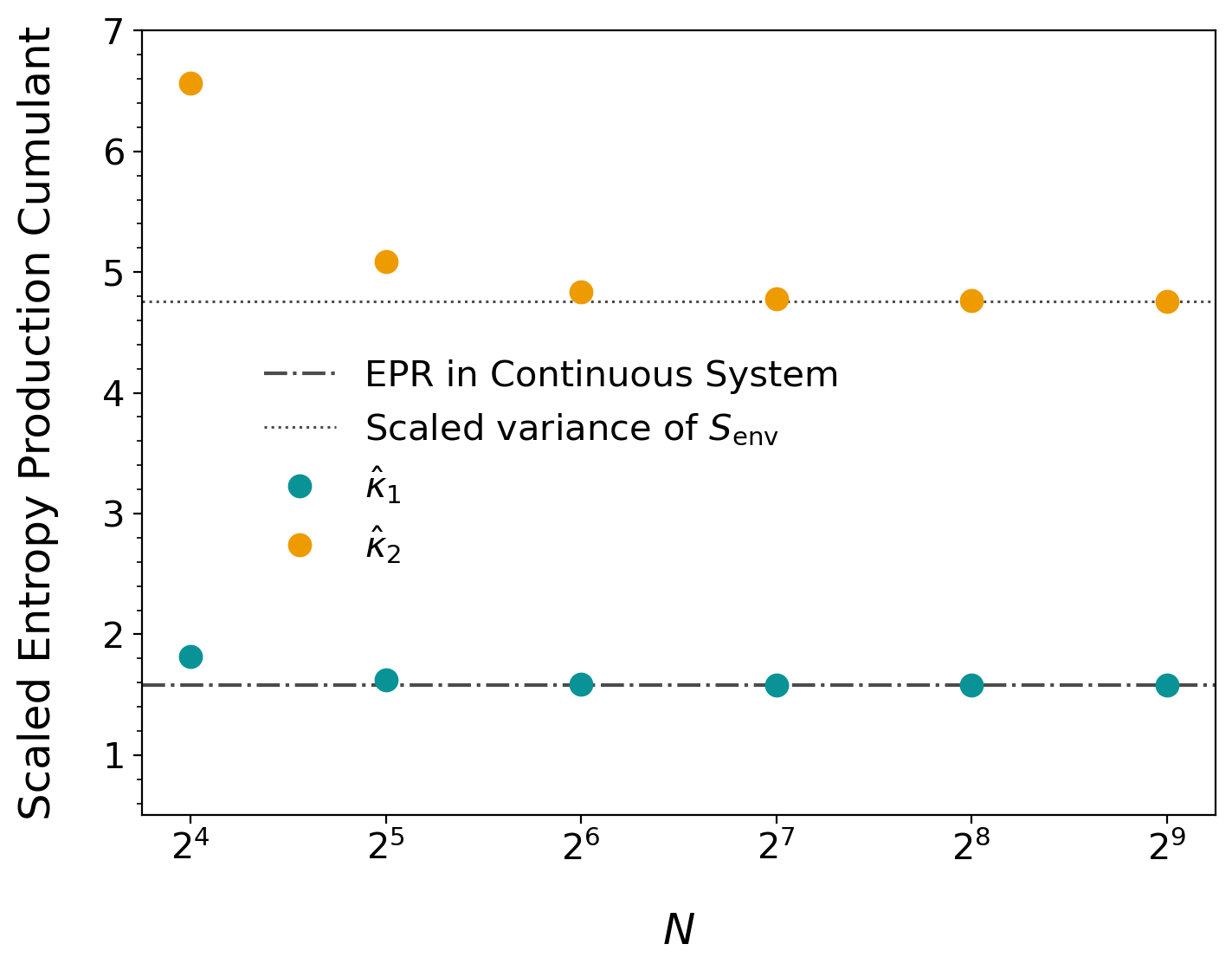}
    \caption{Comparison between scaled cumulants of entropy production in the discrete and the continuous run-and-tumble model. The dashed line is the mean entropy production rate (EPR) given analytically in Ref. \cite{razin2020entropy}. The dotted line is the variance of environmental entropy production calculated from the Langevin simulations of the continuous run-and-tumble model on 1D box within $[-5,5]$ with velocity of the particle $v=1.0$, diffusion coefficient $D=0.5$, and switching rate $r=1.0$. The points are analytically calculated scaled cumulants of environmental entropy production in the discrete run-and-tumble model with scaling of transition rates given by (\ref{discrete-continuous-scaling-relation}). } 
    \label{fig:discrete-continuous-comp}
\end{figure}

Figure~\ref{fig:discrete-continuous-comp} shows the first two scaled cumulants of the entropy production for various system sizes, using the scaling in Eq.~\eqref{discrete-continuous-scaling-relation}, against their value for the continuous system. In particular, the mean entropy production rate has been computed analytically in \cite{razin2020entropy} while we compute the variance of $S_{\rm env}$ in Eq. \eqref{def-entropy-continuous} from the Langevin simulations. The convergence to the continuous case as $N$ increases can be clearly appreciated. Furthermore, due to this convergence, we note that our assumption of local detailed balance in the discrete case is justified a-posteriori.

Unlike the mean entropy production rate, to the best of our knowledge, there have been no theoretical considerations into calculating the variance of entropy production of the run-and-tumble model in continuous space. We have shown that we can compute it using our method under appropriate scaling, and its value converges to what is observed in the continuous system. Similar procedure can also be performed for any moment of the entropy production, but the computation of the third moment in discrete-state system already scales as $\mathcal O (N^3)$, making its computation intensive for a large number of states. 

\section{Summary} \label{sec:summary}
In summary, first we have presented a graphical method to compute the exact moments of entropy production for any discrete-state Markovian system. Employing this method, we have shown that the first and second cumulants scale linearly with time in the long-time limit.

Then, we have applied the developed framework to predict the cumulants of the environmental entropy production in the discrete run-and-tumble model at stationarity, finding non-zero mean, variance and skewness. Additionally, increasing the system size, the environmental entropy production exhibits a remarkable non-Gaussian behavior, highlighting the potential relevance of higher moments when studying the fluctuations of discrete-state systems. Finally, we have performed the continuum limit on the proposed model, finding the correct scaling of the rates with the number of nodes. Within this description, we computed the cumulants of the environmental entropy production for a Langevin run-and-tumble model. We found striking agreement between our predictions, numerical simulations, and a theoretical result previously obtained only for the mean~\cite{razin2020entropy}.

The graphical method presented here can be straightforwardly extended to analyze the moments of currents of any discrete-state systems (and also their continuous counterparts). Our findings suggest that cumulants other than the first two might be relevant in quantifying out-of-equilibrium fluctuations. In principle, one could try to estimate the full probability density function (pdf) of the entropy production including more than the first two moments, using a Maximum Entropy Principle. This task is usually computationally expensive even with only the first three cumulants. Hence, a smarter approach to move from cumulants to an estimation of the PDF would be an interesting topic for future investigations.

Apart from the run-and-tumble model, other common active particle models in the literature are active Brownian particles \cite{romanczuk2012active}, and active Ornstein-Uhlenbeck particles \cite{martin2021statistical}.
Recently, general principles for the entropy production of these active systems have been investigated \cite{frydel2022intuitive}, also in the underdamped regime \cite{frydel2022entropy}. While  results indicated by our work and Ref.~\cite{lebowitz1999gallavotti} give ideas about long time scaling of the cumulants, the exact details about the nature and scaling of entropy production cumulants of various active particle models is another possible line of future investigation. 

\section*{ACKNOWLEDGMENTS}
P.P. acknowledges the support from the University of Padova through the Ph.D. fellowship within ``Bando Dottorati di Ricerca''. D.G. is supported by the Deutsche Forschungsgemeinschaft (DFG, German Research Foundation) - Projekt-nummer 163436311 - SFB 910. The authors thank the referees for relevant and insightful comments, which helped improve the clarity of the manuscript.

\appendix
\section{Third order jump correlation} \label{appendix:third-jump-correlation}
In order to compute the third cumulant, we require all the third order correlations for the number of jumps between states. In this section, following the graphical method, we show some of the terms that arise in this computation. From Fig.~\ref{fig:graphical-method-third-moment}, we choose three orderings of different kinds, a) $k>k'>k''$ where all three $k$-s are different, b) $k>k'=k''$ where only two of the $k$-s are different, and c) $k=k'=k''$ where all three $k$-s are equal.

For $k>k'>k''$, the contribution to third order correlation is
\begin{align}
    \langle n_{m \ell}n_{m' \ell'}n_{m'',\ell''} \rangle_{\Gamma_{\rm MC}} =& \sum_{k=0}^{M-1}\sum_{k'=0}^{k}\sum_{k''=0}^{k'} A_{m \ell}\nonumber \\
    &\times \mathcal P(\ell,k\Delta t|m',(k'+1)\Delta t)~A_{m' \ell'} \nonumber \\
    &\times \mathcal P (\ell',k'\Delta t| m'',(k''+1)\Delta t)\nonumber \\
    &\times A_{m'' \ell''}~P(\ell'',k''\Delta t). 
    \label{eqn:third-jump-correlation-k>k'>k''}
\end{align}

Similarly for $k>k'=k''$,
\begin{align}
    \langle n_{m \ell}n_{m' \ell'}n_{m'',\ell''} \rangle_{\Gamma_{\rm MC}} =& \sum_{k=0}^{M-1}\sum_{k'=0}^k A_{m \ell}\nonumber \\
    &\times \mathcal P(\ell,k\Delta t|m',(k'+1)\Delta t)~\delta_{m',m''}\nonumber \\
    &\times \delta_{\ell',\ell''}~A_{m' \ell'}~P(\ell',k'\Delta t),
\end{align}
and for $k=k'=k''$,
\begin{align}
    \langle n_{m \ell}n_{m' \ell'}n_{m'',\ell''} \rangle_{\Gamma_{\rm MC}} =& \sum_{k=0}^{M-1} \delta_{m,m'}~\delta_{\ell,\ell'}~\delta_{m',m''}~\delta_{\ell',\ell''}\nonumber \\
    &\times A_{m \ell}~P(\ell,k\Delta t) \nonumber \\
    =& \langle n_{m,\ell}\rangle_{\Gamma_{\rm MC}}~\delta_{m,m'}~\delta_{\ell,\ell'}~\delta_{m',m''}~\delta_{\ell',\ell''}.
\end{align}
The contributions from $k<k'=k''$, and $k>k'=k''$ can be written in terms of the second order jump correlations and the first moment of jumps, i.e., 
\begin{align}
    \langle n_{m \ell}n_{m' \ell'}n_{m'',\ell''} \rangle_{\Gamma_{\rm MC}} =~& \langle n_{m \ell}n_{m' \ell'} \rangle_{\Gamma_{\rm MC}}~ \delta_{m',m''}~\delta_{\ell',\ell''} \nonumber \\
    &- \langle n_{m,\ell} \rangle_{\Gamma_{\rm MC}}~\delta_{m',m''}~\delta_{\ell',\ell''} \nonumber \\
    &\times ~ \delta_{m',m''} ~\delta_{\ell',\ell''}
\end{align}
Conversely, starting from a stationary state, $P^{\rm st}$, and taking the limit $\Delta t \to 0$ to recover the master equation formalism, the term arising from Eq. (\ref{eqn:third-jump-correlation-k>k'>k''}) gives the following integral:
\begin{align}
    \langle n_{m \ell}n_{m' \ell'}n_{m'',\ell''} \rangle_T=& W_{ml}~W_{m' \ell'}~W_{m''\ell''}~P^{\rm st}(\ell'') \nonumber \\
    &\times \int_0^T dt\int_0^t dt' ~\mathcal P(\ell,t|m',t')~\nonumber \\ 
    &\times \int_0^{t'}dt'' \mathcal P(\ell',t'|m'',t'')
    \label{eqn:third-jump-correlation-time-integral-unsolved}
\end{align}
Using the eigenvector expansion for the transition probability $\mathcal P(i,t|i_0,t_0)$ (see Eq.~\eqref{exp-P}),
the integrals on the right-hand side of Eq.~\eqref{eqn:third-jump-correlation-time-integral-unsolved} give $\mathcal{T}_{\mathfrak{j}_1, \mathfrak{j}_2}$, i.e, Eq.~\eqref{third-moment-time-integral}.


\begin{thebibliography}{100}

\bibitem{prigogine1971biological}
Ilya Prigogine and Gregoire Nicolis.
\newblock Biological order, structure and instabilities1.
\newblock {\em Quarterly reviews of biophysics}, 4(2-3):107--148, 1971.

\bibitem{Schrodinger1944}
Erwin Schr\"{o}dinger.
\newblock {\em What is Life? The Physical Aspect of the Living Cell}.
\newblock Cambridge University Press, 1944.

\bibitem{fang2019nonequilibrium}
Xiaona Fang, Karsten Kruse, Ting Lu, and Jin Wang.
\newblock Nonequilibrium physics in biology.
\newblock {\em Reviews of Modern Physics}, 91(4):045004, 2019.

\bibitem{ritort2008nonequilibrium}
Felix Ritort.
\newblock Nonequilibrium fluctuations in small systems: From physics to
  biology.
\newblock {\em Advances in chemical physics}, 137:31, 2008.

\bibitem{bustamante2005nonequilibrium}
Carlos Bustamante, Jan Liphardt, and Felix Ritort.
\newblock The nonequilibrium thermodynamics of small systems.
\newblock {\em arXiv preprint cond-mat/0511629}, 2005.

\bibitem{fodor2016far}
{\'E}tienne Fodor, Cesare Nardini, Michael~E Cates, Julien Tailleur, Paolo
  Visco, and Fr{\'e}d{\'e}ric Van~Wijland.
\newblock How far from equilibrium is active matter?
\newblock {\em Physical review letters}, 117(3):038103, 2016.

\bibitem{li2019quantifying}
Junang Li, Jordan~M Horowitz, Todd~R Gingrich, and Nikta Fakhri.
\newblock Quantifying dissipation using fluctuating currents.
\newblock {\em Nature communications}, 10(1):1--9, 2019.

\bibitem{seifert2012stochastic}
Udo Seifert.
\newblock Stochastic thermodynamics, fluctuation theorems and molecular
  machines.
\newblock {\em Reports on progress in physics}, 75(12):126001, 2012.

\bibitem{diana2014mutual}
Giovanni Diana and Massimiliano Esposito.
\newblock Mutual entropy production in bipartite systems.
\newblock {\em Journal of Statistical Mechanics: Theory and Experiment},
  2014(4):P04010, 2014.

\bibitem{busiello2017entropy}
Daniel~M Busiello, Jorge Hidalgo, and Amos Maritan.
\newblock Entropy production in systems with random transition rates close to
  equilibrium.
\newblock {\em Physical Review E}, 96(6):062110, 2017.

\bibitem{busiello2019entropy}
Daniel~M Busiello, Jorge Hidalgo, and Amos Maritan.
\newblock Entropy production for coarse-grained dynamics.
\newblock {\em New Journal of Physics}, 21(7):073004, 2019.

\bibitem{busiello2019entropyB}
Daniel~M Busiello and Amos Maritan.
\newblock Entropy production in master equations and fokker--planck equations:
  facing the coarse-graining and recovering the information loss.
\newblock {\em Journal of Statistical Mechanics: Theory and Experiment},
  2019(10):104013, 2019.

\bibitem{van2003stationary}
R~Van~Zon and EGD Cohen.
\newblock Stationary and transient work-fluctuation theorems for a dragged
  brownian particle.
\newblock {\em Physical Review E}, 67(4):046102, 2003.

\bibitem{douarche2006work}
Fr{\'e}d{\'e}ric Douarche, Sylvain Joubaud, Nicolas~B Garnier, Artyom
  Petrosyan, and Sergio Ciliberto.
\newblock Work fluctuation theorems for harmonic oscillators.
\newblock {\em Physical review letters}, 97(14):140603, 2006.

\bibitem{sabhapandit2012heat}
Sanjib Sabhapandit.
\newblock Heat and work fluctuations for a harmonic oscillator.
\newblock {\em Physical Review E}, 85(2):021108, 2012.

\bibitem{wang2002experimental}
GM~Wang, Edith~M Sevick, Emil Mittag, Debra~J Searles, and Denis~J Evans.
\newblock Experimental demonstration of violations of the second law of
  thermodynamics for small systems and short time scales.
\newblock {\em Physical Review Letters}, 89(5):050601, 2002.

\bibitem{ciliberto2013heat}
Sergio Ciliberto, Alberto Imparato, Antoine Naert, and Marius Tanase.
\newblock Heat flux and entropy produced by thermal fluctuations.
\newblock {\em Physical review letters}, 110(18):180601, 2013.

\bibitem{gupta2018partial}
Deepak Gupta and Sanjib Sabhapandit.
\newblock Partial entropy production in heat transport.
\newblock {\em Journal of Statistical Mechanics: Theory and Experiment},
  2018(6):063203, 2018.

\bibitem{shreshtha2019thermodynamic}
Mayank Shreshtha and Rosemary~J Harris.
\newblock Thermodynamic uncertainty for run-and-tumble--type processes.
\newblock {\em EPL (Europhysics Letters)}, 126(4):40007, 2019.

\bibitem{Van-Zon-heat}
R.~van Zon and E.~G.~D. Cohen.
\newblock Extension of the fluctuation theorem.
\newblock {\em Phys. Rev. Lett.}, 91:110601, Sep 2003.

\bibitem{visco2006work}
Paolo Visco.
\newblock Work fluctuations for a brownian particle between two thermostats.
\newblock {\em Journal of Statistical Mechanics: Theory and Experiment},
  2006(06):P06006, 2006.

\bibitem{kundu2011large}
Anupam Kundu, Sanjib Sabhapandit, and Abhishek Dhar.
\newblock Large deviations of heat flow in harmonic chains.
\newblock {\em Journal of Statistical Mechanics: Theory and Experiment},
  2011(03):P03007, 2011.

\bibitem{martinez2016brownian}
Ignacio~A Mart{\'\i}nez, {\'E}dgar Rold{\'a}n, Luis Dinis, Dmitri Petrov,
  Juan~MR Parrondo, and Ra{\'u}l~A Rica.
\newblock Brownian carnot engine.
\newblock {\em Nature physics}, 12(1):67--70, 2016.

\bibitem{van2012efficiency}
Christian Van~den Broeck, Niraj Kumar, and Katja Lindenberg.
\newblock Efficiency of isothermal molecular machines at maximum power.
\newblock {\em Physical review letters}, 108(21):210602, 2012.

\bibitem{verley2014unlikely}
Gatien Verley, Massimiliano Esposito, Tim Willaert, and Christian Van~den
  Broeck.
\newblock The unlikely carnot efficiency.
\newblock {\em Nature communications}, 5(1):1--5, 2014.

\bibitem{gupta2017stochastic}
Deepak Gupta and Sanjib Sabhapandit.
\newblock Stochastic efficiency of an isothermal work-to-work converter engine.
\newblock {\em Physical Review E}, 96(4):042130, 2017.

\bibitem{gupta2018exact}
Deepak Gupta.
\newblock Exact distribution for work and stochastic efficiency of an
  isothermal machine.
\newblock {\em Journal of Statistical Mechanics: Theory and Experiment},
  2018(7):073201, 2018.

\bibitem{seifert2005entropy}
Udo Seifert.
\newblock Entropy production along a stochastic trajectory and an integral
  fluctuation theorem.
\newblock {\em Physical review letters}, 95(4):040602, 2005.

\bibitem{sekimoto2010stochastic}
Ken Sekimoto.
\newblock {\em Stochastic energetics}, volume 799.
\newblock Springer, 2010.

\bibitem{kurchan1998fluctuation}
Jorge Kurchan.
\newblock Fluctuation theorem for stochastic dynamics.
\newblock {\em Journal of Physics A: Mathematical and General}, 31(16):3719,
  1998.

\bibitem{crooks1999entropy}
Gavin~E Crooks.
\newblock Entropy production fluctuation theorem and the nonequilibrium work
  relation for free energy differences.
\newblock {\em Physical Review E}, 60(3):2721, 1999.

\bibitem{lebowitz1999gallavotti}
Joel~L Lebowitz and Herbert Spohn.
\newblock A gallavotti--cohen-type symmetry in the large deviation functional
  for stochastic dynamics.
\newblock {\em Journal of Statistical Physics}, 95(1):333--365, 1999.

\bibitem{campisi2009fluctuation}
Michele Campisi, Peter Talkner, and Peter H{\"a}nggi.
\newblock Fluctuation theorem for arbitrary open quantum systems.
\newblock {\em Physical review letters}, 102(21):210401, 2009.

\bibitem{jarzynski1997nonequilibrium}
Christopher Jarzynski.
\newblock Nonequilibrium equality for free energy differences.
\newblock {\em Physical Review Letters}, 78(14):2690, 1997.

\bibitem{crooks1998nonequilibrium}
Gavin~E Crooks.
\newblock Nonequilibrium measurements of free energy differences for
  microscopically reversible markovian systems.
\newblock {\em Journal of Statistical Physics}, 90(5):1481--1487, 1998.

\bibitem{baiesi2013update}
Marco Baiesi and Christian Maes.
\newblock An update on the nonequilibrium linear response.
\newblock {\em New Journal of Physics}, 15(1):013004, 2013.

\bibitem{baiesi2009fluctuations}
Marco Baiesi, Christian Maes, and Bram Wynants.
\newblock Fluctuations and response of nonequilibrium states.
\newblock {\em Physical review letters}, 103(1):010602, 2009.

\bibitem{dechant2018multidimensional}
Andreas Dechant.
\newblock Multidimensional thermodynamic uncertainty relations.
\newblock {\em Journal of Physics A: Mathematical and Theoretical},
  52(3):035001, 2018.

\bibitem{barato2015thermodynamic}
Andre~C Barato and Udo Seifert.
\newblock Thermodynamic uncertainty relation for biomolecular processes.
\newblock {\em Physical review letters}, 114(15):158101, 2015.

\bibitem{gingrich2016dissipation}
Todd~R Gingrich, Jordan~M Horowitz, Nikolay Perunov, and Jeremy~L England.
\newblock Dissipation bounds all steady-state current fluctuations.
\newblock {\em Physical review letters}, 116(12):120601, 2016.

\bibitem{horowitz2020thermodynamic}
Jordan~M Horowitz and Todd~R Gingrich.
\newblock Thermodynamic uncertainty relations constrain non-equilibrium
  fluctuations.
\newblock {\em Nature Physics}, 16(1):15--20, 2020.

\bibitem{manikandan2021quantitative}
Sreekanth~K Manikandan, Subhrokoli Ghosh, Avijit Kundu, Biswajit Das, Vipin
  Agrawal, Dhrubaditya Mitra, Ayan Banerjee, and Supriya Krishnamurthy.
\newblock Quantitative analysis of non-equilibrium systems from short-time
  experimental data.
\newblock {\em Communications Physics}, 4(1):1--10, 2021.

\bibitem{das2022inferring}
Biswajit Das, Sreekanth~K Manikandan, and Ayan Banerjee.
\newblock Inferring entropy production in anharmonic brownian gyrators.
\newblock {\em arXiv preprint arXiv:2204.09283}, 2022.

\bibitem{van2020entropy}
Tan Van~Vu, Yoshihiko Hasegawa, et~al.
\newblock Entropy production estimation with optimal current.
\newblock {\em Physical Review E}, 101(4):042138, 2020.

\bibitem{skinner2021estimating}
Dominic~J Skinner and J{\"o}rn Dunkel.
\newblock Estimating entropy production from waiting time distributions.
\newblock {\em Physical review letters}, 127(19):198101, 2021.

\bibitem{otsubo2020estimating}
Shun Otsubo, Sosuke Ito, Andreas Dechant, and Takahiro Sagawa.
\newblock Estimating entropy production by machine learning of short-time
  fluctuating currents.
\newblock {\em Physical Review E}, 101(6):062106, 2020.

\bibitem{roldan2010estimating}
{\'E}dgar Rold{\'a}n and Juan~MR Parrondo.
\newblock Estimating dissipation from single stationary trajectories.
\newblock {\em Physical review letters}, 105(15):150607, 2010.

\bibitem{otsubo2022estimating}
Shun Otsubo, Sreekanth~K Manikandan, Takahiro Sagawa, and Supriya
  Krishnamurthy.
\newblock Estimating time-dependent entropy production from non-equilibrium
  trajectories.
\newblock {\em Communications Physics}, 5(1):1--10, 2022.

\bibitem{lander2012noninvasive}
Boris Lander, Jakob Mehl, Valentin Blickle, Clemens Bechinger, and Udo Seifert.
\newblock Noninvasive measurement of dissipation in colloidal systems.
\newblock {\em Physical Review E}, 86(3):030401, 2012.

\bibitem{saha2009entropy}
Arnab Saha, Sourabh Lahiri, and AM~Jayannavar.
\newblock Entropy production theorems and some consequences.
\newblock {\em Physical Review E}, 80(1):011117, 2009.

\bibitem{gupta2016fluctuation}
Deepak Gupta and Sanjib Sabhapandit.
\newblock Fluctuation theorem for entropy production of a partial system in the
  weak-coupling limit.
\newblock {\em EPL (Europhysics Letters)}, 115(6):60003, 2016.

\bibitem{gupta2020entropy}
Deepak Gupta and Sanjib Sabhapandit.
\newblock Entropy production for partially observed harmonic systems.
\newblock {\em Journal of Statistical Mechanics: Theory and Experiment},
  2020(1):013204, 2020.

\bibitem{martynec2020entropy}
Thomas Martynec, Sabine~HL Klapp, and Sarah~AM Loos.
\newblock Entropy production at criticality in a nonequilibrium potts model.
\newblock {\em New Journal of Physics}, 22(9):093069, 2020.

\bibitem{tietz2006measurement}
C~Tietz, S~Schuler, T~Speck, U~Seifert, and J~Wrachtrup.
\newblock Measurement of stochastic entropy production.
\newblock {\em Physical review letters}, 97(5):050602, 2006.

\bibitem{speck2007distribution}
Thomas Speck, Valentin Blickle, Clemens Bechinger, and Udo Seifert.
\newblock Distribution of entropy production for a colloidal particle in a
  nonequilibrium steady state.
\newblock {\em EPL (Europhysics Letters)}, 79(3):30002, 2007.

\bibitem{koski2013distribution}
JV~Koski, T~Sagawa, OP~Saira, Y~Yoon, A~Kutvonen, P~Solinas,
  M~M{\"o}tt{\"o}nen, T~Ala-Nissila, and JP~Pekola.
\newblock Distribution of entropy production in a single-electron box.
\newblock {\em Nature Physics}, 9(10):644--648, 2013.

\bibitem{manikandan2022nonmonotonic}
Sreekanth~K. Manikandan, Biswajit Das, Avijit Kundu, Raunak Dey, Ayan Banerjee,
  and Supriya Krishnamurthy.
\newblock Nonmonotonic skewness of currents in nonequilibrium steady states.
\newblock {\em Phys. Rev. Research}, 4:043067, Oct 2022.

\bibitem{ciliberto2017experiments}
Sergio Ciliberto.
\newblock Experiments in stochastic thermodynamics: Short history and
  perspectives.
\newblock {\em Physical Review X}, 7(2):021051, 2017.

\bibitem{seara2018entropy}
Daniel~S Seara, Vikrant Yadav, Ian Linsmeier, A~Pasha Tabatabai, Patrick~W
  Oakes, SM~Tabei, Shiladitya Banerjee, and Michael~P Murrell.
\newblock Entropy production rate is maximized in non-contractile actomyosin.
\newblock {\em Nature communications}, 9(1):1--10, 2018.

\bibitem{ramaswamy2010mechanics}
Sriram Ramaswamy.
\newblock The mechanics and statistics of active matter.
\newblock {\em Annu. Rev. Condens. Matter Phys.}, 1(1):323--345, 2010.

\bibitem{martinez2017colloidal}
Ignacio~A Martinez, {\'E}dgar Rold{\'a}n, Luis Dinis, and Ra{\'u}l~A Rica.
\newblock Colloidal heat engines: a review.
\newblock {\em Soft matter}, 13(1):22--36, 2017.

\bibitem{busiello2021dissipation}
Daniel~M Busiello, Shiling Liang, Francesco Piazza, and Paolo De~Los~Rios.
\newblock Dissipation-driven selection of states in non-equilibrium chemical
  networks.
\newblock {\em Communications Chemistry}, 4(1):1--7, 2021.

\bibitem{dass2021equilibrium}
Avinash~Vicholous Dass, Thomas Georgelin, Frances Westall, Fr{\'e}d{\'e}ric
  Foucher, Paolo De~Los~Rios, Daniel~M Busiello, Shiling Liang, and Francesco
  Piazza.
\newblock Equilibrium and non-equilibrium furanose selection in the ribose
  isomerisation network.
\newblock {\em Nature Communications}, 12(1):1--10, 2021.

\bibitem{manikandan2020inferring}
Sreekanth~K Manikandan, Deepak Gupta, and Supriya Krishnamurthy.
\newblock Inferring entropy production from short experiments.
\newblock {\em Physical review letters}, 124(12):120603, 2020.

\bibitem{speck2005integral}
Thomas Speck and Udo Seifert.
\newblock Integral fluctuation theorem for the housekeeping heat.
\newblock {\em Journal of Physics A: Mathematical and General}, 38(34):L581,
  2005.

\bibitem{dechant2018entropic}
Andreas Dechant and Shin-ichi Sasa.
\newblock Entropic bounds on currents in langevin systems.
\newblock {\em Physical Review E}, 97(6):062101, 2018.

\bibitem{pietzonka2016universal}
Patrick Pietzonka, Andre~C Barato, and Udo Seifert.
\newblock Universal bounds on current fluctuations.
\newblock {\em Physical Review E}, 93(5):052145, 2016.

\bibitem{marchetti2013hydrodynamics}
M~Cristina Marchetti, Jean-Fran{\c{c}}ois Joanny, Sriram Ramaswamy,
  Tanniemola~B Liverpool, Jacques Prost, Madan Rao, and R~Aditi Simha.
\newblock Hydrodynamics of soft active matter.
\newblock {\em Reviews of modern physics}, 85(3):1143, 2013.

\bibitem{ramaswamy2017active}
Sriram Ramaswamy.
\newblock Active matter.
\newblock {\em Journal of Statistical Mechanics: Theory and Experiment},
  2017(5):054002, 2017.

\bibitem{prost2015active}
Jacques Prost, Frank J{\"u}licher, and Jean-Fran{\c{c}}ois Joanny.
\newblock Active gel physics.
\newblock {\em Nature physics}, 11(2):111--117, 2015.

\bibitem{stenhammar2016light}
Joakim Stenhammar, Raphael Wittkowski, Davide Marenduzzo, and Michael~E Cates.
\newblock Light-induced self-assembly of active rectification devices.
\newblock {\em Science advances}, 2(4):e1501850, 2016.

\bibitem{du2019self}
Yunfei Du, Huijun Jiang, and Zhonghuai Hou.
\newblock Self-assembly of active core corona particles into highly ordered and
  self-healing structures.
\newblock {\em The Journal of Chemical Physics}, 151(15):154904, 2019.

\bibitem{stenhammar2015activity}
Joakim Stenhammar, Raphael Wittkowski, Davide Marenduzzo, and Michael~E Cates.
\newblock Activity-induced phase separation and self-assembly in mixtures of
  active and passive particles.
\newblock {\em Physical review letters}, 114(1):018301, 2015.

\bibitem{pu2017reentrant}
Mingfeng Pu, Huijun Jiang, and Zhonghuai Hou.
\newblock Reentrant phase separation behavior of active particles with
  anisotropic janus interaction.
\newblock {\em Soft matter}, 13(22):4112--4121, 2017.

\bibitem{fily2012athermal}
Yaouen Fily and M~Cristina Marchetti.
\newblock Athermal phase separation of self-propelled particles with no
  alignment.
\newblock {\em Physical review letters}, 108(23):235702, 2012.

\bibitem{mandal2017entropy}
Dibyendu Mandal, Katherine Klymko, and Michael~R DeWeese.
\newblock Entropy production and fluctuation theorems for active matter.
\newblock {\em Physical review letters}, 119(25):258001, 2017.

\bibitem{caprini2019entropy}
Lorenzo Caprini, Umberto Marini~Bettolo Marconi, Andrea Puglisi, and Angelo
  Vulpiani.
\newblock The entropy production of ornstein--uhlenbeck active particles: a
  path integral method for correlations.
\newblock {\em Journal of Statistical Mechanics: Theory and Experiment},
  2019(5):053203, 2019.

\bibitem{martin2021statistical}
David Martin, J{\'e}r{\'e}my O'Byrne, Michael~E Cates, {\'E}tienne Fodor,
  Cesare Nardini, Julien Tailleur, and Fr{\'e}d{\'e}ric van Wijland.
\newblock Statistical mechanics of active ornstein-uhlenbeck particles.
\newblock {\em Physical Review E}, 103(3):032607, 2021.

\bibitem{speck2016stochastic}
Thomas Speck.
\newblock Stochastic thermodynamics for active matter.
\newblock {\em EPL (Europhysics Letters)}, 114(3):30006, 2016.

\bibitem{szamel2019stochastic}
Grzegorz Szamel.
\newblock Stochastic thermodynamics for self-propelled particles.
\newblock {\em Physical Review E}, 100(5):050603, 2019.

\bibitem{chaki2019effects}
Subhasish Chaki and Rajarshi Chakrabarti.
\newblock Effects of active fluctuations on energetics of a colloidal particle:
  Superdiffusion, dissipation and entropy production.
\newblock {\em Physica A: Statistical Mechanics and its Applications},
  530:121574, 2019.

\bibitem{nardini2017entropy}
Cesare Nardini, {\'E}tienne Fodor, Elsen Tjhung, Fr{\'e}d{\'e}ric Van~Wijland,
  Julien Tailleur, and Michael~E Cates.
\newblock Entropy production in field theories without time-reversal symmetry:
  quantifying the non-equilibrium character of active matter.
\newblock {\em Physical Review X}, 7(2):021007, 2017.

\bibitem{shankar2018hidden}
Suraj Shankar and M~Cristina Marchetti.
\newblock Hidden entropy production and work fluctuations in an ideal active
  gas.
\newblock {\em Physical Review E}, 98(2):020604, 2018.

\bibitem{cao2019effective}
Zhiyu Cao, Jie Su, Huijun Jiang, and Zhonghuai Hou.
\newblock Effective entropy production and thermodynamic uncertainty relation
  of active brownian particles.
\newblock {\em arXiv preprint arXiv:1907.11459}, 2019.

\bibitem{skinner2021improved}
Dominic~J Skinner and J{\"o}rn Dunkel.
\newblock Improved bounds on entropy production in living systems.
\newblock {\em Proceedings of the National Academy of Sciences}, 118(18), 2021.

\bibitem{gupta2021heat}
Deepak Gupta and David~A Sivak.
\newblock Heat fluctuations in a harmonic chain of active particles.
\newblock {\em Physical Review E}, 104(2):024605, 2021.

\bibitem{falasco2016exact}
Gianmaria Falasco, Richard Pfaller, Andreas~P Bregulla, Frank Cichos, and Klaus
  Kroy.
\newblock Exact symmetries in the velocity fluctuations of a hot brownian
  swimmer.
\newblock {\em Physical Review E}, 94(3):030602, 2016.

\bibitem{berg2004coli}
Howard~C Berg.
\newblock {\em E. coli in Motion}.
\newblock Springer, 2004.

\bibitem{elgeti2015run}
Jens Elgeti and Gerhard Gompper.
\newblock Run-and-tumble dynamics of self-propelled particles in confinement.
\newblock {\em EPL (Europhysics Letters)}, 109(5):58003, 2015.

\bibitem{demaerel2018active}
Thibaut Demaerel and Christian Maes.
\newblock Active processes in one dimension.
\newblock {\em Physical Review E}, 97(3):032604, 2018.

\bibitem{malakar2018steady}
Kanaya Malakar, V~Jemseena, Anupam Kundu, K~Vijay Kumar, Sanjib Sabhapandit,
  Satya~N Majumdar, S~Redner, and Abhishek Dhar.
\newblock Steady state, relaxation and first-passage properties of a
  run-and-tumble particle in one-dimension.
\newblock {\em Journal of Statistical Mechanics: Theory and Experiment},
  2018(4):043215, 2018.

\bibitem{rosenau1993random}
Philip Rosenau.
\newblock Random walker and the telegrapher’s equation: A paradigm of a
  generalized hydrodynamics.
\newblock {\em Physical Review E}, 48(2):R655, 1993.

\bibitem{weiss2002some}
George~H Weiss.
\newblock Some applications of persistent random walks and the telegrapher's
  equation.
\newblock {\em Physica A: Statistical Mechanics and its Applications},
  311(3-4):381--410, 2002.

\bibitem{evans2018run}
Martin~R Evans and Satya~N Majumdar.
\newblock Run and tumble particle under resetting: a renewal approach.
\newblock {\em Journal of Physics A: Mathematical and Theoretical},
  51(47):475003, 2018.

\bibitem{dhar2019run}
Abhishek Dhar, Anupam Kundu, Satya~N Majumdar, Sanjib Sabhapandit, and
  Gr{\'e}gory Schehr.
\newblock Run-and-tumble particle in one-dimensional confining potentials:
  steady-state, relaxation, and first-passage properties.
\newblock {\em Physical Review E}, 99(3):032132, 2019.

\bibitem{angelani2015run}
Luca Angelani.
\newblock Run-and-tumble particles, telegrapher’s equation and absorption
  problems with partially reflecting boundaries.
\newblock {\em Journal of Physics A: Mathematical and Theoretical},
  48(49):495003, 2015.

\bibitem{razin2020entropy}
Nitzan Razin.
\newblock Entropy production of an active particle in a box.
\newblock {\em Physical Review E}, 102(3):030103, 2020.

\bibitem{van1992stochastic}
Nicolaas~Godfried Van~Kampen.
\newblock {\em Stochastic processes in physics and chemistry}, volume~1.
\newblock Elsevier, 1992.

\bibitem{esposito2010entropy}
Massimiliano Esposito, Katja Lindenberg, and Christian Van~den Broeck.
\newblock Entropy production as correlation between system and reservoir.
\newblock {\em New Journal of Physics}, 12(1):013013, 2010.

\bibitem{schuster2013nonequilibrium}
Heinz~Georg Schuster.
\newblock {\em Nonequilibrium statistical physics of small systems: Fluctuation
  relations and beyond}.
\newblock John Wiley \& Sons, 2013.

\bibitem{katz1983phase}
Sheldon Katz, Joel~L Lebowitz, and H~Spohn.
\newblock Phase transitions in stationary nonequilibrium states of model
  lattice systems.
\newblock {\em Physical Review B}, 28(3):1655, 1983.


\bibitem{maes2021local}
Christian Maes.
\newblock Local detailed balance.
\newblock {\em SciPost Physics Lecture Notes}, page 032, 2021.


\bibitem{li2021steady}
Yuting~I Li and Michael~E Cates.
\newblock Steady state entropy production rate for scalar langevin field
  theories.
\newblock {\em Journal of Statistical Mechanics: Theory and Experiment},
  2021(1):013211, 2021.

\bibitem{busiello2020entropy}
Daniel~M Busiello, Deepak Gupta, and Amos Maritan.
\newblock Entropy production in systems with unidirectional transitions.
\newblock {\em Physical Review Research}, 2(2):023011, 2020.

\bibitem{gardiner1985handbook}
Crispin~W Gardiner et~al.
\newblock {\em Handbook of stochastic methods}, volume~3.
\newblock springer Berlin, 1985.

\bibitem{busiello2022hyperaccurate}
Daniel~M Busiello and Carlos Fiore.
\newblock Hyperaccurate bounds in discrete-state markovian systems.
\newblock {\em arXiv preprint arXiv:2205.00294}, 2022.

\bibitem{romanczuk2012active}
Pawel Romanczuk, Markus B{\"a}r, Werner Ebeling, Benjamin Lindner, and Lutz
  Schimansky-Geier.
\newblock Active brownian particles.
\newblock {\em The European Physical Journal Special Topics}, 202(1):1--162,
  2012.

\bibitem{frydel2022intuitive}
Derek Frydel.
\newblock Intuitive view of entropy production of ideal run-and-tumble
  particles.
\newblock {\em Physical Review E}, 105(3):034113, 2022.

\bibitem{frydel2022entropy}
Derek Frydel.
\newblock Entropy production of active particles in underdamped regime.
\newblock {\em arXiv preprint arXiv:2211.02082}, 2022.

\end{thebibliography}

\end{document}